\documentclass[manuscript]{aa} 
\usepackage{graphicx}
\usepackage{txfonts}
\usepackage[]{natbib}
\usepackage{multirow}
\usepackage[colorlinks=true,linkcolor=blue,citecolor=blue]{hyperref}
\def\deg{$^{\circ}$~}
\def\degb{^{\circ}}

\begin{document}

 \title{Morphology of the very inclined debris disk around HD\,32297\thanks{Based on observations collected at the European Southern
Observatory, Chile, ESO (385.C-0476(B))}}

 \author{
 	A. Boccaletti\inst{1},  
 	J.-C. Augereau\inst{2},  
	A.-M. Lagrange\inst{2},  
	J. Milli\inst{2},  
	P. Baudoz\inst{1}, 
	D. Mawet\inst{3},
	D. Mouillet\inst{2},
	J. Lebreton   \inst{2}, \and
	A.-L. Maire  \inst{1}
 }
 
 \offprints{A. Boccaletti, \email{anthony.boccaletti@obspm.fr} }

 \institute{LESIA, Observatoire de Paris, CNRS, University Pierre et Marie Curie Paris 6 and University Denis Diderot Paris 7, 5 place Jules Janssen, 92195 Meudon, France
             \and
             Institut de Plan\'etologie et d'Astrophysique de Grenoble, Universit\'e Joseph Fourier, CNRS, BP 53, 38041 Grenoble, France
             \and
           European Southern Observatory, Casilla 19001, Santiago 19, Chile
                  }

   \date{Received XXX; Accepted XXX}

  \keywords{Stars: early-type -- Stars: individual (HD\,32297) -- Techniques: image processing -- Techniques: high angular resolution}

\authorrunning{A. Boccaletti et al.}
\titlerunning{Imaging of HD\,32297}

 \abstract
{Direct imaging of circumstellar disks at high angular resolution is mandatory to provide morphological information that bring constraints on their properties, in particular the spatial distribution of dust. This challenging objective was for a long time in most cases, only within the realm of space telescopes from the visible to the infrared. New techniques combining observing strategy and data processing now allow very high contrast imaging with 8-m class ground-based telescopes ($10^{-4}$ to $10^{-5}$ at $\sim$1") and complement space telescopes while improving angular resolution at near infrared wavelengths. }
{We carried out a program at the VLT with NACO to image known debris disks  with higher angular resolution in the near IR than ever before in order to study morphological properties and ultimately to detect signpost of planets.}
{The observing method makes use of advanced techniques: Adaptive Optics, Coronagraphy and Differential Imaging, a combination designed to directly image exoplanets with the upcoming generation of "planet finders" like GPI (Gemini Planet Imager) and SPHERE (Spectro-Polarimetric High contrast Exoplanet REsearch). Applied to extended objects like circumstellar disks, the method is still successful but produces significant biases in terms of photometry and morphology. We developed a new model-matching procedure to correct for these biases and hence to bring constraints on the morphology of debris disks.}
{From our program, we present new images of the disk around the star HD\,32297 obtained in the H (1.6$\mu m$)  and Ks (2.2$\mu m$) bands with an unprecedented angular resolution ($\sim$65 mas). 
The images show an inclined thin disk detected at separations larger than $0.5-0.6"$. The modeling stage confirms a very high inclination ($i=88\degb$) and the presence of an inner cavity inside $r_0\approx 110$\,AU.  We also found that the spine (line of maximum intensity along the midplane)  of the disk is curved and we attributed this feature to a large anisotropic scattering factor ($g \approx 0.5$, valid for an non-edge on disk).}
{Our modeling procedure is relevant to interpret images of circumstellar disks observed with Angular Differential Imaging. 
It allows both to reduce the biases and to provide an estimation of disk parameters.
}
 \maketitle

%

\section{Introduction}

\begin{table*}[t]
\caption{Log of observations summarizing the parameters: the filter used, the time of observation, the integration time of a single frame (DIT), the number of frames per data cube (NDIT), the number of cubes  (Nexp), the seeing variation, the mean correlation time $\tau_0$,
the amplitude of the parallactic angle variation and the observation time on target.}
\begin{center}
\begin{tabular}{ccccccccccc} \hline\hline
Date			& 	Filter &	UT start/end		&	DIT (s)	&	NDIT		&	Nexp	& seeing (")	& $\tau_0$ (ms)	& 	Parall. amplitude ($^\circ$) & { on target (s)}	\\	\hline
2010.09.25	&	H	&	07:23:40 / 08:45:15	&	2 / 3		& 	15 / 10	&	{ 68/13}		& 0.70 - 1.28	& 1.7			&	24.81  & { 1710} \\
2010.09.25	&	Ks	&	09:04:08 / 10:00:57	&	2		&	15		&	57		& 0.59 - 0.88	& 2.0			&	24.18 & { 2430} \\ \hline 
\end{tabular}
\end{center}
\label{tab:log}
\end{table*}

Many stars have photometric IR excess attributed to the presence of circumstellar dust. Among the circumstellar disk evolution, 
the debris disk phase is the period where most of the primordial gas { has been} dissipated or accreted onto already formed giant planets. The dust content responsible for the IR excesses is thought to be the result of regular mutual collisions between asteroid-like objects and/or evaporation of cometary bodies \citep{Wyatt2008, Krivov2010}. This activity, presumably triggered by unseen planets, is continuously replenishing the system with observable dust grains that, in turn, trace out the presence of a planetary system. The recent detection by direct imaging of planets in circumstellar disks around main sequence A-type stars like $\beta$ Pictoris \citep{Lagrange2010} and HR 8799 \citep{Marois2008} or the undefined point source around Fomalhaut \citep{Kalas2008, Janson2012} has clearly revealed the suspected intimate relation between { planet} formation and debris disks structures \citep{Ozernoy2000, Wyatt2003, Reche2008}. 
Moreover, the most prominent structures in these { specific} disks (warp, annulus, gap, offset) can be accounted for by the mass and orbital properties of { these} associated planets. { Similar features are seen in other debris disks where planets are not yet imaged.} 
Many disks are thought to have internal cavities attributed to dust sweeping by 
planets. From a face-on to an edge-on geometry these structures are more or less easily identified. Therefore, a morphological analysis is mandatory to study disk evolution, planetary formation and { the likely} presence of planets. 

HD\,32297 is a young A5 star identified with an IR excess ($L_{IR}/L_*=0.0027$) by IRAS and later resolved as a { inclined} circumstellar debris disk by \citet{Schneider2005} in the near IR with NICMOS on HST.
Observations in the visible (R band) were obtained by \citet{Kalas2005} and reveal a circumstellar nebulosity at distances larger than 560\,AU indicating a broad scattered-light region extending to 1680\,AU.
The brightness asymmetries and a distorted morphology suggests the interaction with the interstellar medium. 
HST data obtained at 1.1$\mu$\,m probe the inner region below $\sim$400\,AU and as close as 34\,AU from the star (corresponding to respectively 3.3" and 0.3"). \citet{Schneider2005}  derived a Position Angle ($PA$) of $47.6\pm1.0$\deg and an inclination of about 10\deg from edge-on. A surface brightness asymmetry is also observed in the South West (SW) side of the disk which is about twice brighter than the North East (NE) side at a projected angular separation of about 0.3-0.4".
Observations at mid-IR wavelengths (11.7 and 18.3$\mu$\,m) by \citet{Moerchen2007} indicate a dust depletion closer than $\sim$70\,AU suggestive of a inner cavity like the one found in HR\,4796 \citep{Schneider1999} or HD\,61005 \citep{Buenzli2010} but with a less steep decrease of the density. However, this radius is not well constrained in the mid IR as the equivalent resolution corresponds to 35 to 50\,AU at the HD\,32297 distance.
Additional colors were obtained by \citet{Debes2009} at 1.60 and 2.05$\mu$\,m again with NICMOS which confirms the disk orientation ($PA=46.3\pm1.3$\deg at NE and  $49.0\pm2.0$\deg at SW). The surface brightness profiles confirm the asymmetries and reveal a slope break at 90\,AU (NE)  and 110\,AU (SW). If { similar to some} other disks like $\beta$ Pictoris \citep{Augereau2001}  and AU Mic \citep{Augereau2006} this break indicates the boundary of a planetesimals disk, it is marginally compatible with the value (70\,AU) derived by \citet{Moerchen2007} but the measurements may not be directly comparable to that of \citet{Schneider2005}. Also, it is shown that the disk midplane is curved at separations larger than 1.5" (170\,AU), possibly a result of the interaction with the interstellar medium. This probably impacts the $PA$ measurement. 
More recently, \citet{Mawet2009} achieved very high contrast imaging of the disk with a small (1.6\,m) well corrected sub-aperture at the Palomar in Ks and confirmed the colors and brightness asymmetry, but the lower angular resolution did not allow a detailed morphological study. 

At a distance of $113 \pm 12$\,pc \citep{Perryman1997}, the disk of HD\,32297 is of particular interest to probe the regions where planets would have formed in the 10-AU scale providing high angular resolution and high contrast are achieved simultaneously. In this paper, we present new diffraction limited images of the HD\,32297 debris disk in the near IR (1.6 and 2.2 $\mu$m) using an 8-m class telescope { with adaptive optics}, therefore with a significant improvement with respect to former observations (a factor of $\sim$3 { for the angular resolution} at the same wavelengths). 
We note that similar observations obtained at the Keck telescope in the same near IR bands are presented by \citet{Esposito2012}.
Section \ref{sec:observ} presents the results of the observation. In section \ref{sec:morpho} we analyze the reliability of a particular structure observed in the image. Section \ref{sec:photom} provides the Surface Brightness profiles of the disk before biases are taken into account. The core of this paper is presented in the section \ref{sec:model} where we attempt to identify the parameters of the disk. In section \ref{sec:photomcorrec} the Surface Brightness corrected for the observational bias is derived. Finally, we present constraints on the presence of point sources within the disk. 

\section{Observation and data reduction}
\label{sec:observ}
\subsection{Observations}
\label{sec:observobserv}

The star HD\,32297 (A5, V=8.13, H=7.62, K=7.59, near-IR values are from 2MASS) was observed
with NACO, the near IR AO-assisted camera at the VLT \citep{Rousset2003, Lenzen2003}, on Sept. 25th, 2010 UT using a dedicated strategy to allow for high contrast imaging. We combined two modes offered in NACO: the four-quadrant phase mask coronagraphs \citep[FQPMs,][]{Rouan2000} to attenuate the on-axis star and the pupil tracking mode to allow for Angular Differential Imaging \citep[ADI,][]{Marois2006}. We used the broad-band filters H and Ks for which the FQPMs are optimized \citep{Boccaletti2004} and the S27 camera providing a sampling of $\sim$0.027" per pixel. Coronagraphic images are obtained with a field stop of $8\times8$" and  $13\times13$" in respectively  H and Ks and an undersized pupil stop of 90\% the full pupil diameter. 
Phase masks already provided interesting results on both extended objects \citep{Riaud2006, Boccaletti2009, Mawet2009} and point sources as companion to stars \citep{Boccaletti2008, Serabyn2010, Mawet2011}. 

The log of observations is shown in Table \ref{tab:log}. The integration time of individual FQPM exposures (DIT) was chosen to fill about 80\% of the dynamical range of the detector (including the attenuation by the coronagraph), and the duration of a template (DIT $\times$ NDIT) do not exceed 30s so as to correct for a systematic drift correlated with the parallactic angle variations in pupil tracking mode. Therefore, the star is re-aligned with the coronagraphic mask every 30s when needed (the drift increases closer to the meridian).
At the end of the sequence, a few sky frames were acquired around the target star.
As for calibration purpose, we obtained out-of-mask images of the star itself with a neutral density filter (but without the undersized stop) to determine the point-spread function (PSF) which is intended to normalize coronagraphic images and hence to calculate surface brightness or contrast curves.
Also, since the phase masks are engraved on a glass substrate, their transmission map is measured with a flat illumination provided by the internal lamp of NACO.
Seeing conditions were good to medium: $0.72\pm0.06$" in Ks and $0.97\pm0.12$" in H. { The AO correction provides a FWHM (Full Width at Half Maximum) of $\sim$65mas in both filters while theoretical values for an 8-m class telescope at the longest wavelengths of the H and Ks bands are 46 and 59 mas respectively. Therefore, images are nearly diffraction-limited in Ks as opposed to H.}

In pupil tracking, while the field rotates, the smearing of off-axis objects is faster at the edge of the field. However, for a few seconds of integration per frame, the smearing is totally negligible within our field of view.

\subsection{Data reduction and results}
\label{sec:reduction}

\begin{figure*}[ht]
\centerline{
\includegraphics[width=6cm]{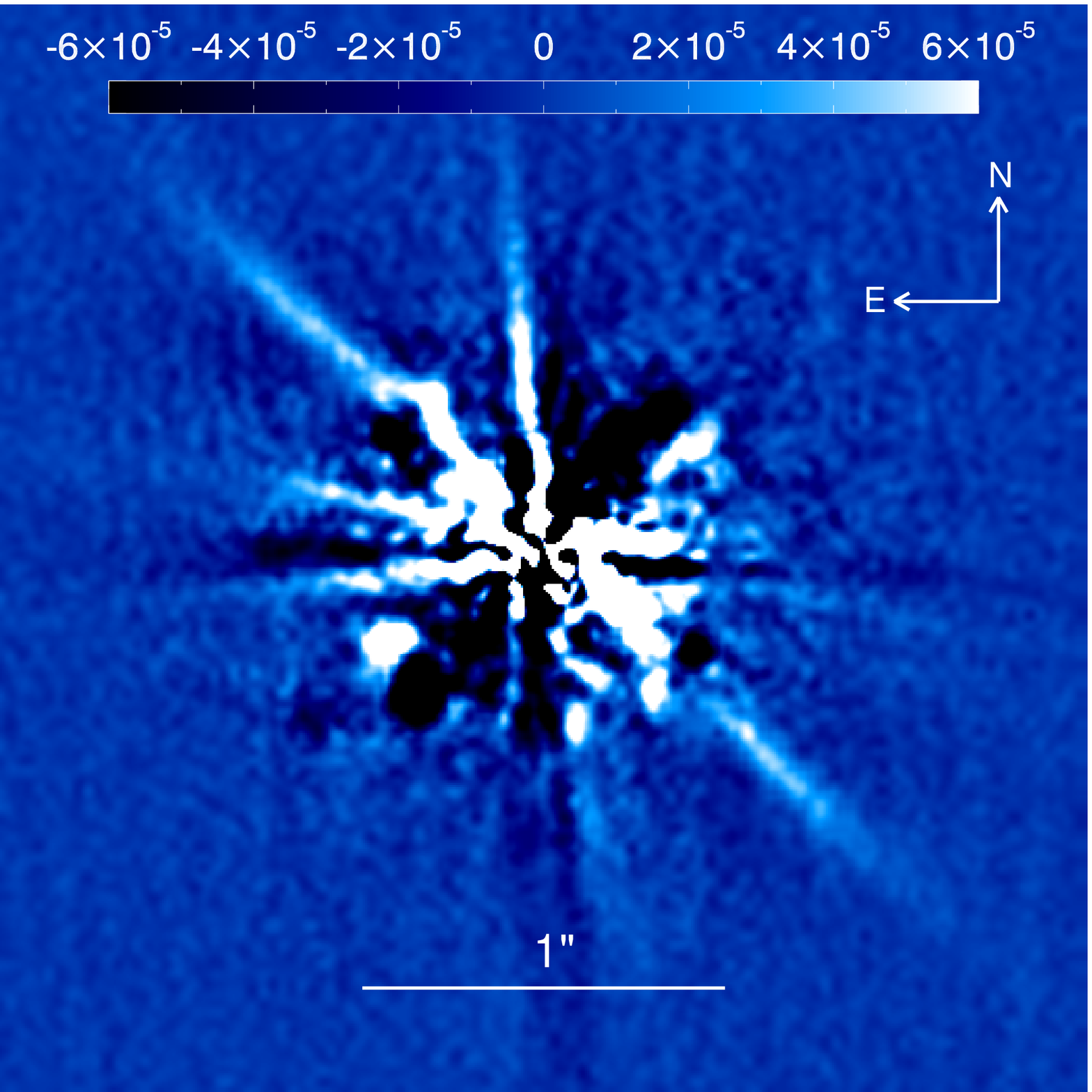}
\includegraphics[width=6cm]{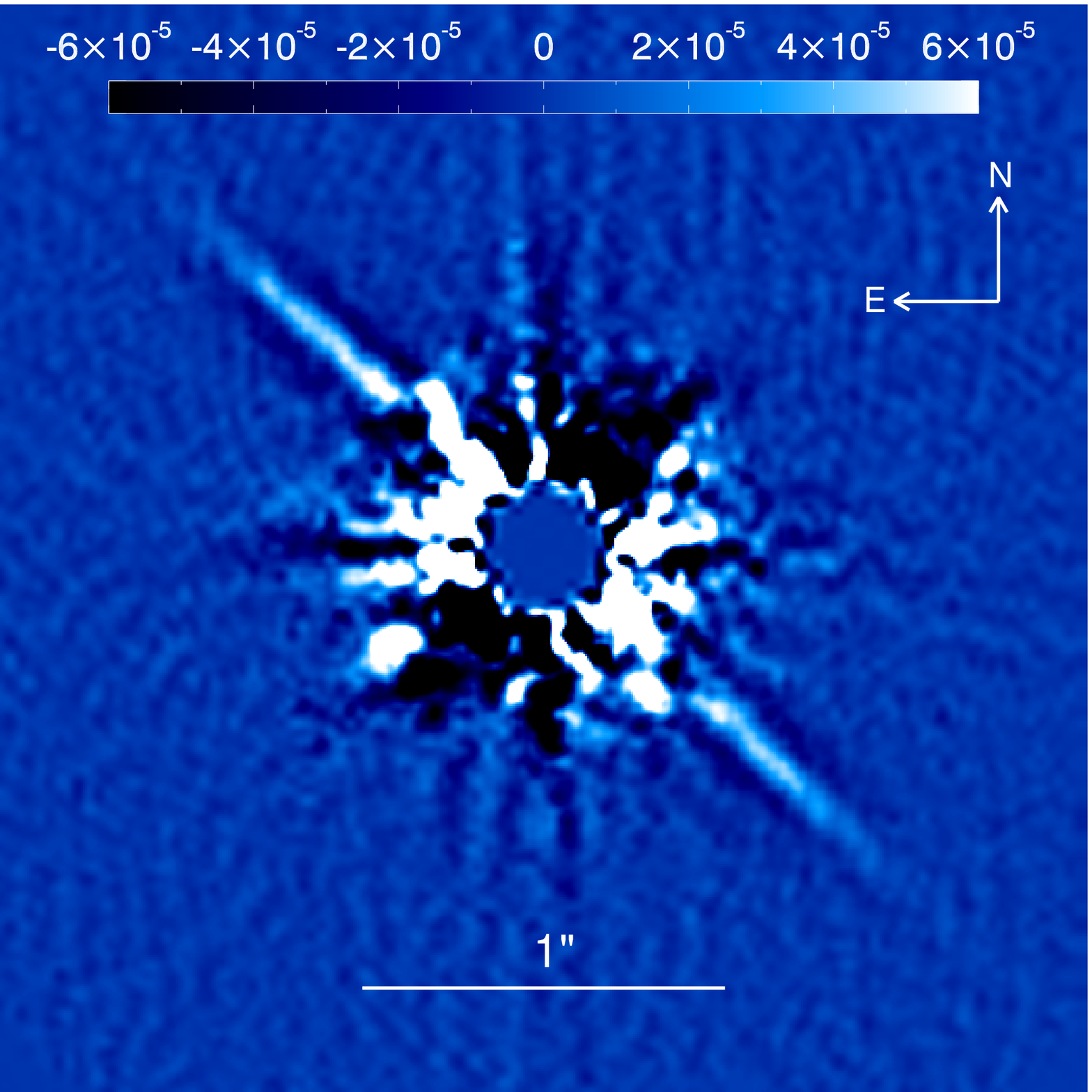}
\includegraphics[width=6cm]{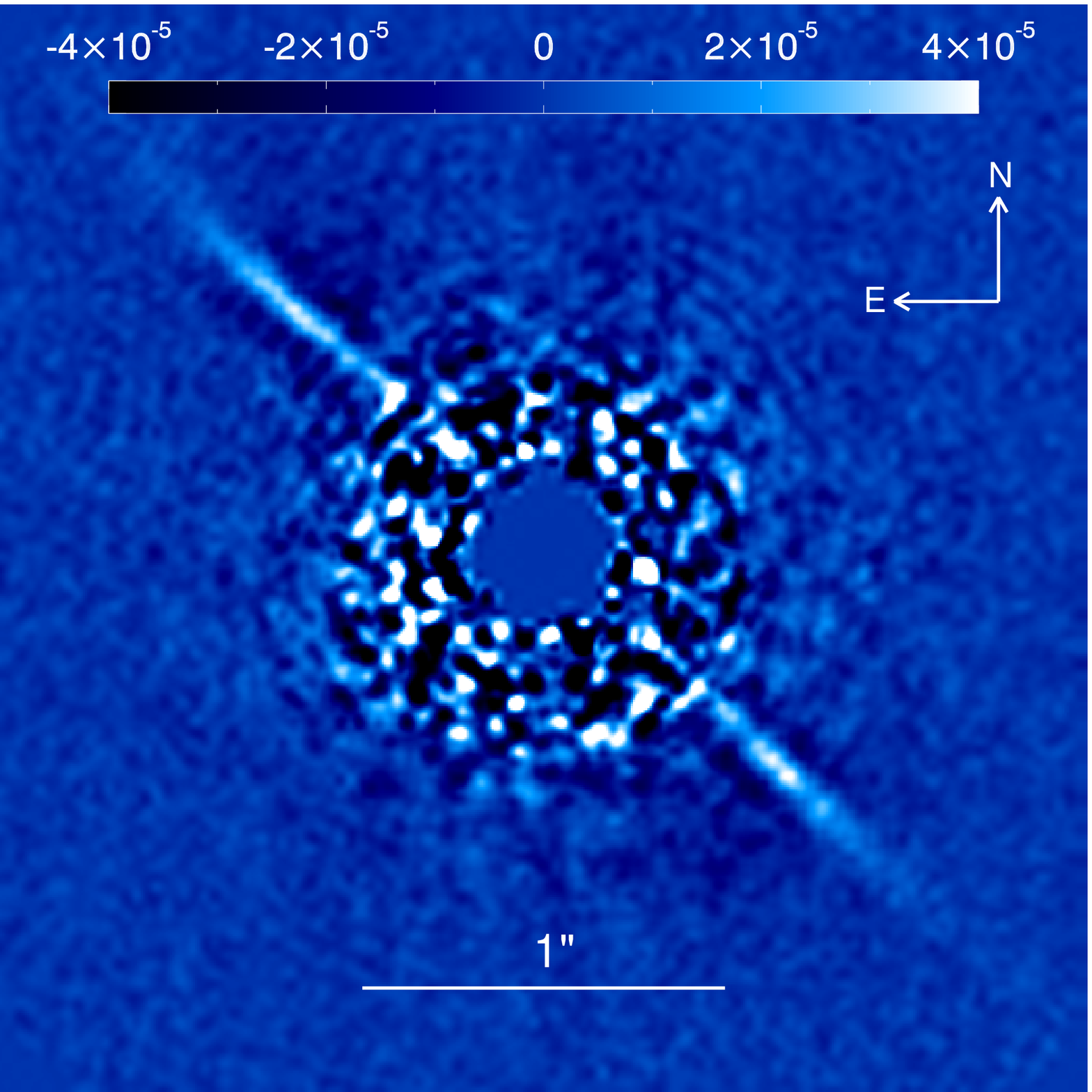}
}
\centerline{
\includegraphics[width=6cm]{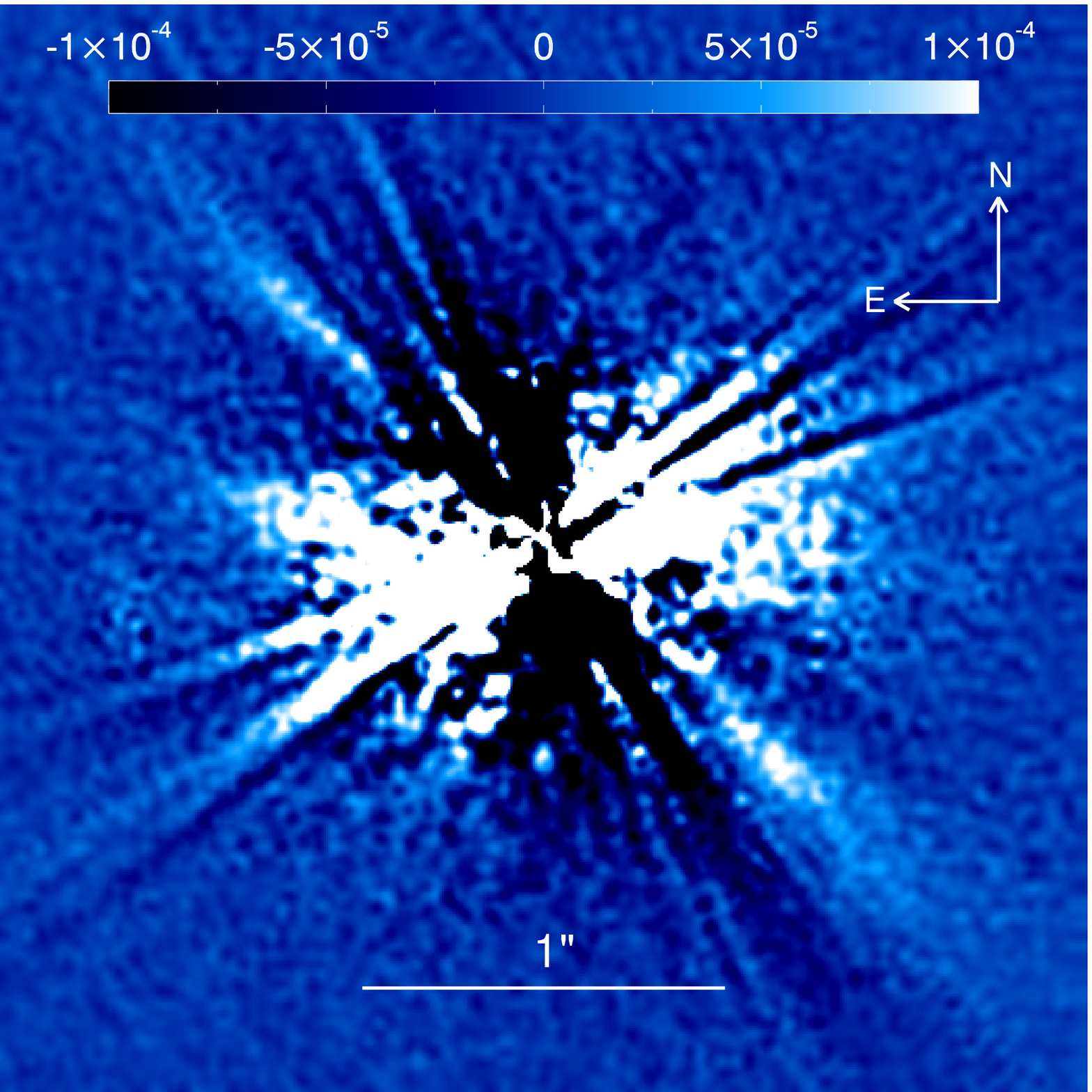}
\includegraphics[width=6cm]{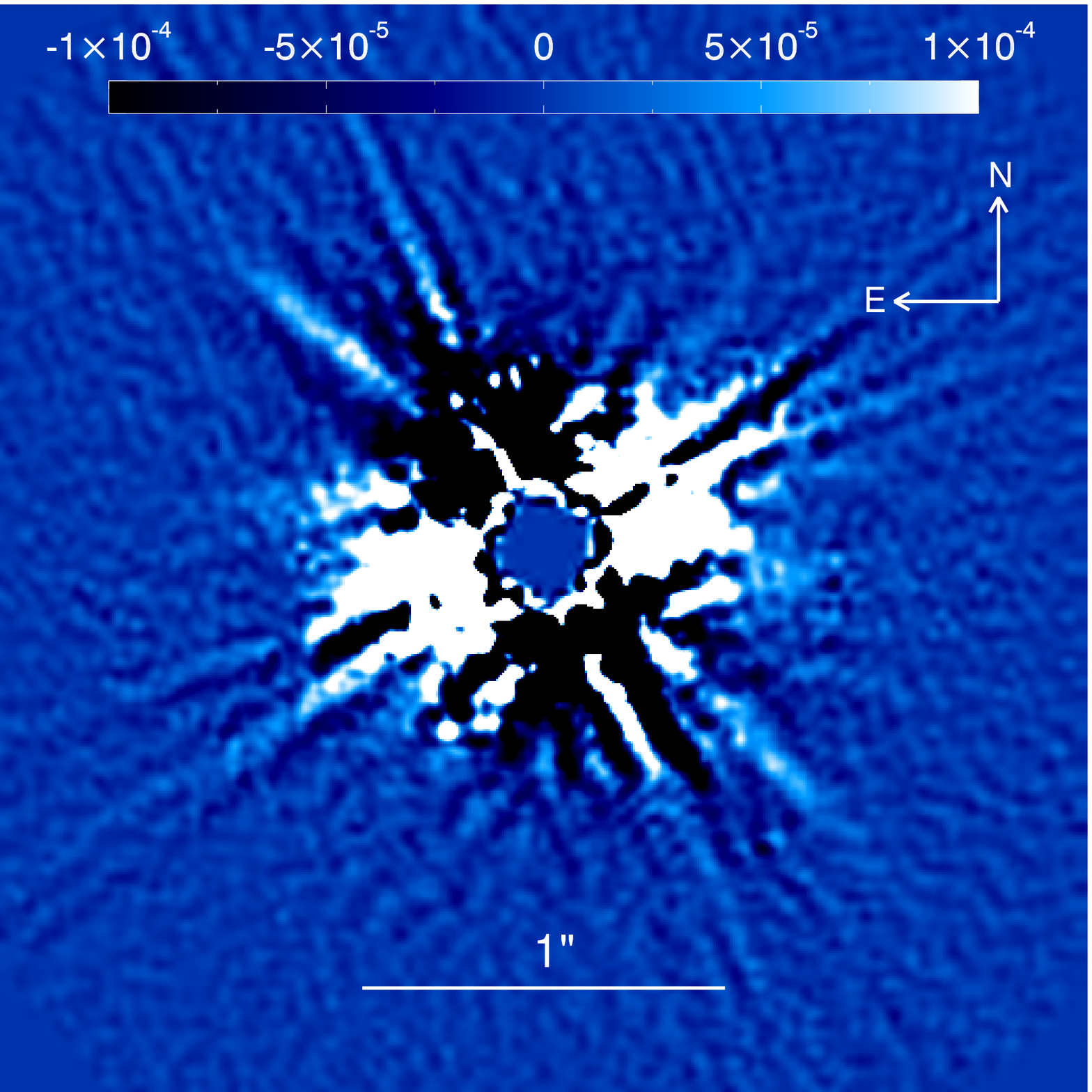}
\includegraphics[width=6cm]{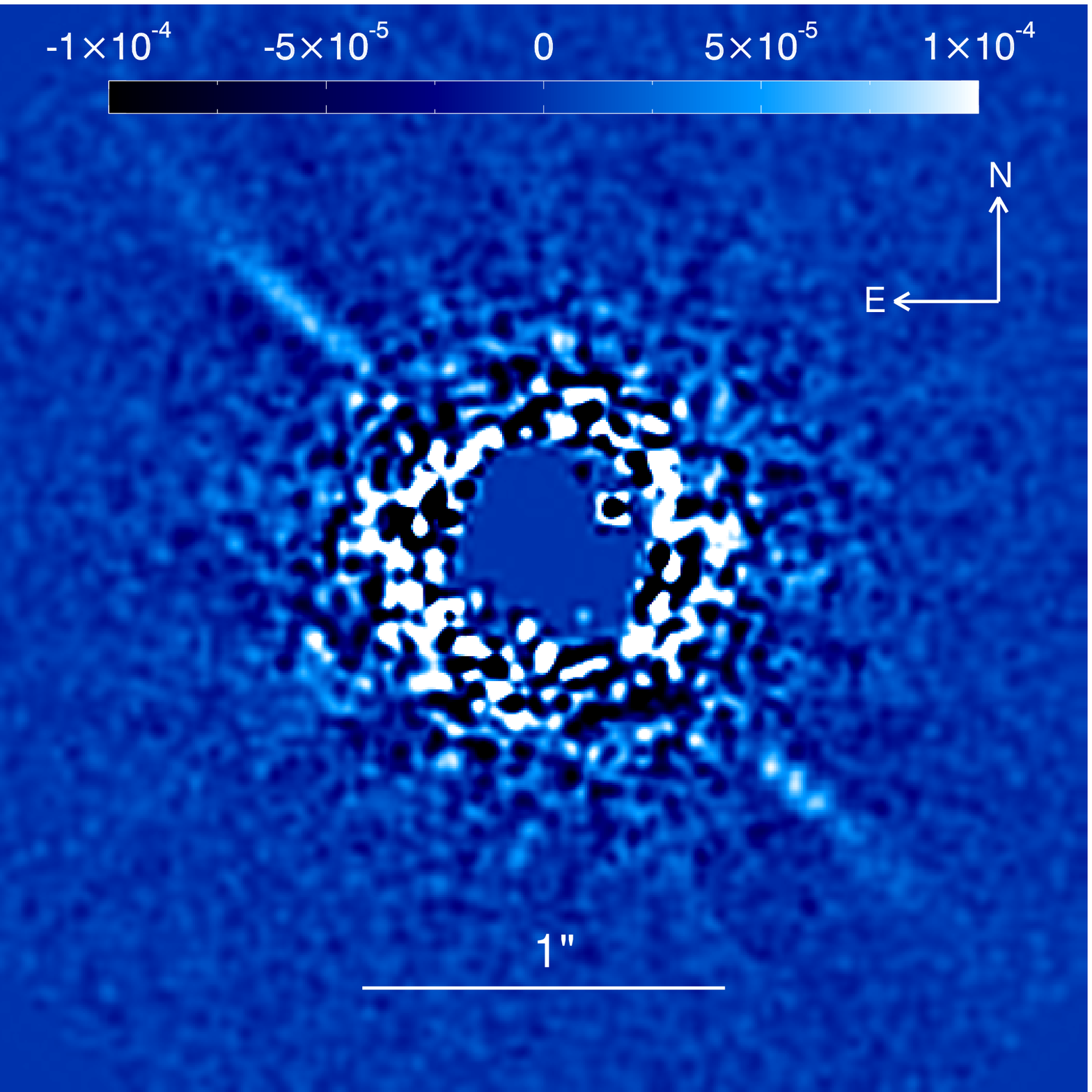}
}
\caption{Final images of the disk around HD\,32297 in the Ks (top) and H (bottom) bands for the cADI, rADI, and LOCI (left to right) algorithms (see section \ref{sec:reduction} for acronyms). The field of view of each sub-image is 3$\times$3". { The color bar indicates the contrast with respect to the star per pixel (not applicable to the display-saturated central residuals).} 
}
\label{fig:results}
\end{figure*}

The data were reduced with our own pipeline. First, we applied bad pixel removal and Flat Field correction to every frame. Although quite faint here, we averaged sky frames and subtract a background map (including the dark) as it reduces the offset.
Given the large amount of coronagraphic images and to save computing time we binned the data cubes to reduce their size. We adopt a binning step of 5 frames (averaging)
while keeping the ADI-induced smearing still negligible.
A median sky frame is subtracted to every binned frames. Then, we rejected the frames of worse quality with a flux criteria (open AO loop for instance) and obtained data cubes of 129 (75\% of data) and 158 (69\% of data) frames for respectively the Ks and H band data, representing a total integration time of 1290 and 1706 seconds respectively.
The frame recentering is a non linear problem, because the star moves behind the coronagraph so the intensity and shape of the residual pattern vary significantly. It is therefore not possible to apply standard centroiding algorithm and the issues are different than in saturated imaging \citep{Lagrange2012}. In particular, the FQPM creates four peaks of unequal intensities which are difficult to exploit to provide an astrometric signal since phase aberrations are also quite large and perturb the coronagraphic image.
We tested several procedures, focusing on the fitting of the halo. 
The best result is obtained when the individual frames are thresholded to 10\% of the image maximum and a Moffat profile is fitted so that to give more weight to some pixels in the stellar halo with respect to those that are closer to the star. We achieved this way a sub-pixel accuracy for centering but we have not reliable mean to estimate a value.
Finally, we used the background image of the FQPM alone to apply a correction of its own Flat Field to every re-centered frame. 

Following the nomenclature described in \citet{Lagrange2012} we processed the data with several ADI algorithms (which differ in the way to build a reference frame): 
\begin{itemize}
\item[-]{classical ADI \citep[cADI,][]{Marois2006} which subtracts the median of the datacube to each frame.}
\item[-]{radial ADI \citep[rADI,][]{Marois2006, Lafreniere2007} for which a set of reference frames to be subtracted is determined separately on each annulus. The algorithm is controlled by the separation criterion which defines in a deterministic way, together with the amount of field rotation, the time lapse between one frame and its reference frames.
}
\item[-]{Locally Optimized Combination of Images \citep[LOCI,][]{Lafreniere2007} is more sophisticated since it defines optimization zones (with a particular geometry) where the reference frame to be subtracted (in a zone smaller than the optimization one) is estimated as a linear combination of all frames where the field object has moved by the desired quantity (separation criterion $N_{\delta}$). Optimization zones are larger than the subtraction zones to limit the subtraction of a potential point source. Several parameters control the LOCI algorithm but the most relevant here are the separation criterion and the size of the optimization zone.}
\end{itemize}
After each algorithm, the frames are derotated to put the North vertical and a median is taken over to provide the final image.

As circumstellar disks are extended objects, these techniques are not optimized { since they favor the detection of unresolved sources and apply in different independent zones in the image breaking the continuous shape of a disk.
Still, they can be very efficient to emphasize regions with strong gradients of similar spatial scales as point sources (rings for instance). 
In the past few years several impressive results were obtained using ADI and revealed the very detailed structures within some debris disks \citep{Buenzli2010, Thalmann2011, Lagrange2012, Lagrange2012b}. 
Self-subtraction is an important issue and is caused by the contribution of the disk itself to the building of the reference image to be subtracted to the data cube.}
Because of the field rotation, the inner regions { of a circumstellar disks} are more affected by self-subtraction in the ADI processing as they cover smaller angular sectors. Moreover, the self-subtraction { of the disk increases} as the algorithm ability to suppress speckles improves (LOCI being more aggressive than cADI for instance). This is particularly important in our data since the rotation amplitude is less than 25$\degb$. A detailed analysis of the self-subtraction arising when imaging disks with ADI will be presented in Milli et al. (submitted).

ADI algorithms can produce a wealth of various images across which both the stellar residuals and the signal of the object will be modulated. 
The adjustment of the parameters relies on the balance between self-subtraction and the separation from the star at which the disk is detected. This qualitative analysis leads us to set the separation criterion for both rADI and LOCI to 1.0 and 1.5$\times$FWHM for respectively the Ks and H bands, where FWHM=65 mas in both filters. 
The radial width is 2$\times$FWHM for both the LOCI subtraction zones and rADI rings. As for LOCI, we adopt the geometry described in \citet{Lafreniere2007} where the zones are distributed centro-symmetrically. 
Subtraction zones have a radial to azimuth size factor of unity. 
The optimization zones start at the same radius than the subtraction zones but extend at larger radii with a total area of $N_A=300\times$PSF cores. To build the reference image in each zone, all the available frames were used and the coefficients are calculated with a singular value decomposition. 

\begin{figure}[th]
\centerline{\includegraphics[width=8cm]{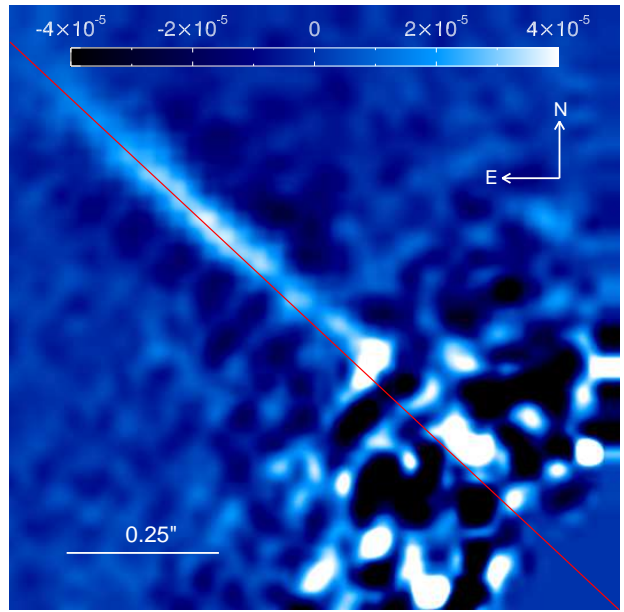}
}
\caption{ Upper left 1$\times$1" quadrant of the Ks LOCI image which emphasizes the deviation with respect to the midplane (red line, $PA = 47.4\degb$). The star is at the lower-right corner. The color bar indicates the contrast with respect to the star per pixel (not applicable to the display-saturated central residuals). }
\label{fig:zoom}
\end{figure}

\begin{figure}[th]
\centerline{\includegraphics[width=8cm]{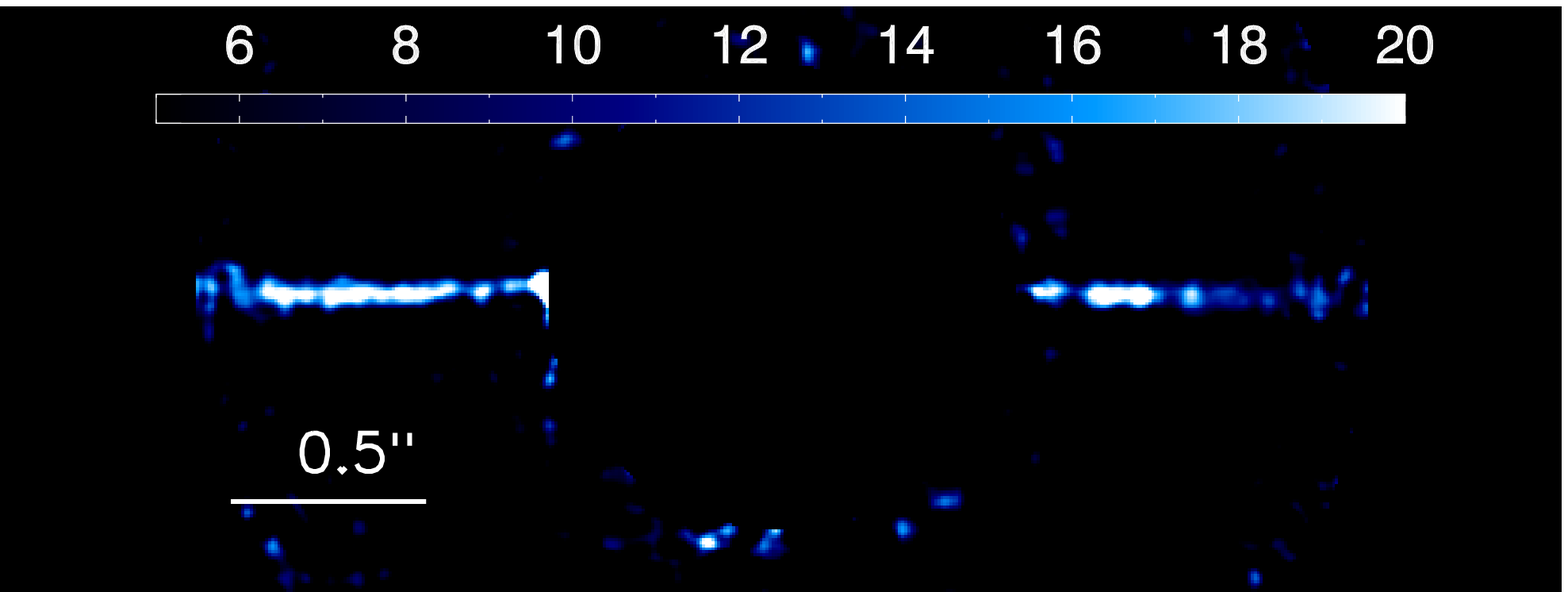}}
\centerline{\includegraphics[width=8cm]{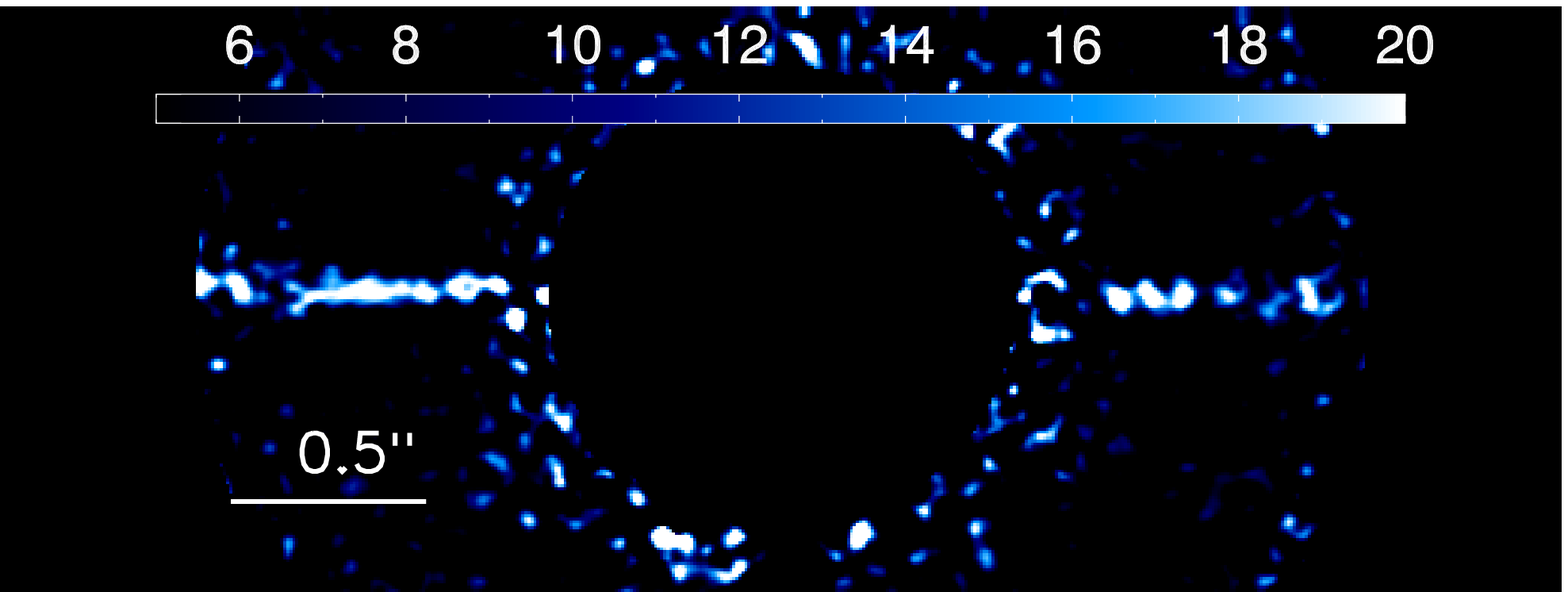}}
\caption{$SNR$ map obtained from LOCI images in Ks (top) and H (bottom) for $SNR>5\sigma$ per resolution element. The field of view is 4"x1.5" and the disk is derotated ($PA=47.4\degb$) to put the disk horizontal. The outer ($>1.5"$) and inner ($<0.6"$) regions are excluded. The color bar indicates the $SNR$ per resolution elements.}
\label{fig:snrmap}
\end{figure}

A first look at the resulting images shown in Fig. \ref{fig:results} indicates that 1/ no information is retrieved at separation closer than 0.5-0.6" since stellar residuals still dominate the image (for instance, the cADI images show strong waffle pattern as an imprint of the AO system), 
2/ the H band data are of lower quality than in Ks (as expected from AO performance, see section \ref{sec:observobserv}, and seeing variations, see Table \ref{tab:log}) but LOCI efficiently reduces the stellar residuals at similar radii in both bands, 3/ the disk has about the same thickness through all algorithms especially in the Ks band which favors a thin and highly inclined disk, 4/ the H and Ks morphologies are very similar. 
The structure of stellar residuals changes according to the algorithm, and differs notably from the well-known speckled pattern, especially owing to the azimuthal averaging caused by the de-rotation of the frames.
A closer inspection shows that although the disk is quite linear it features a small deviation from the midplane which is more {emphasized} in LOCI images to the NE side (especially in Ks but also in H). 
Figure \ref{fig:zoom} shows the position of the midplane (red line) with respect to the spine of the disk. The deviation is evidenced in the middle of this image and spans several resolution elements  (from $\sim$0.6" to $\sim$1.0").
The main objective of the paper is to find a disk morphological model which can account for this deviation.
In addition, we present signal to noise ratio maps in Fig. \ref{fig:snrmap}. The noise level was estimated azimuthally on LOCI images (both H and Ks) which are smoothed with a gaussian (FWHM = 65ms) so as to derive the SNR per resolution element. We achieve detection levels of about 10 to 20$\sigma$ along the disk midplane. We warn that the noise estimation can be biased especially in the speckle-dominated regime close to the star as explained in \citet{Marois2008b}. In the case of Fig. \ref{fig:snrmap} this occurs at distances closer than $0.5 \sim 0.6"$ although a few speckles still appear  further out as significant pattern in the H band $SNR$ map. Finally, we note that the deviation from the midplane is also visible in the SNR maps. 


\section{Morphological analysis}
\label{sec:morpho}
\subsection{Position Angle determination}

First, we measured the $PA$ of the disk to check its consistency with the values reported previously by \citet{Schneider2005} and \citet{Debes2009}, and with the goal to improve the accuracy owing to the higher angular resolution and the benefit of the ADI processing.
For that, we followed the approach presented in \citet{Boccaletti2009} and more recently in \citet{Lagrange2012} for the case of $\beta$ Pictoris,  as it is nearly edge-on as the one around HD\,32297. The image is de-rotated by an angle corresponding to a rough estimation of the PA (guess) so that the disk is about horizontal with respect to pixels. Vertical fitting with a gaussian profile along the columns yields for each separation the position of the spine { (trace of the maximum surface brightness along the midplane)} at a sub-pixel precision. 
Then, from the position of the spine as a function of radius (smoothed over 4 pixels $\approx$2 FWHM), we estimated the slope of the spine in four regions carefully chosen to reduce the influence of stellar residuals and background noise. The measurements were made in the regions 25-40, 25-50, 30-40 and 30-50  pixels (corresponding to a range of 0.675 - 1.35") on each side of the disk { together} (assuming the disk crosses the star). 

The process is repeated for several angles in a range of 2\deg sampled at 1/100\deg around the guessed $PA$. Finally, the spine slope as a function of this angle shows a steep minimum which defines unambiguously (at a precision of $0.2\degb$) the $PA$ of the disk. We made the same analysis for all the images in Fig. \ref{fig:results} and averaged the values. 
The error bar is calculated from the dispersion of the values to which we added quadratically the Celestial North uncertainty \citep[0.2\deg according to][]{Lagrange2012}.
 We found $PA = 47.4\pm0.3$\deg in the Ks band and $PA=47.0\pm0.3$\deg in the H band which agree within error bars but we adopt the former value since Ks data are of much better quality.
These numbers compare adequately to previous reports but achieve better precision. However, we did not observe a $PA$ variation of about 3$\degb$ between the two sides of the disk unlike \citet{Debes2009} but we perform our measurement at closer separations than in HST { so the studies are not directly comparable.

\begin{figure}[lt]
\centerline{\includegraphics[width=8.5cm]{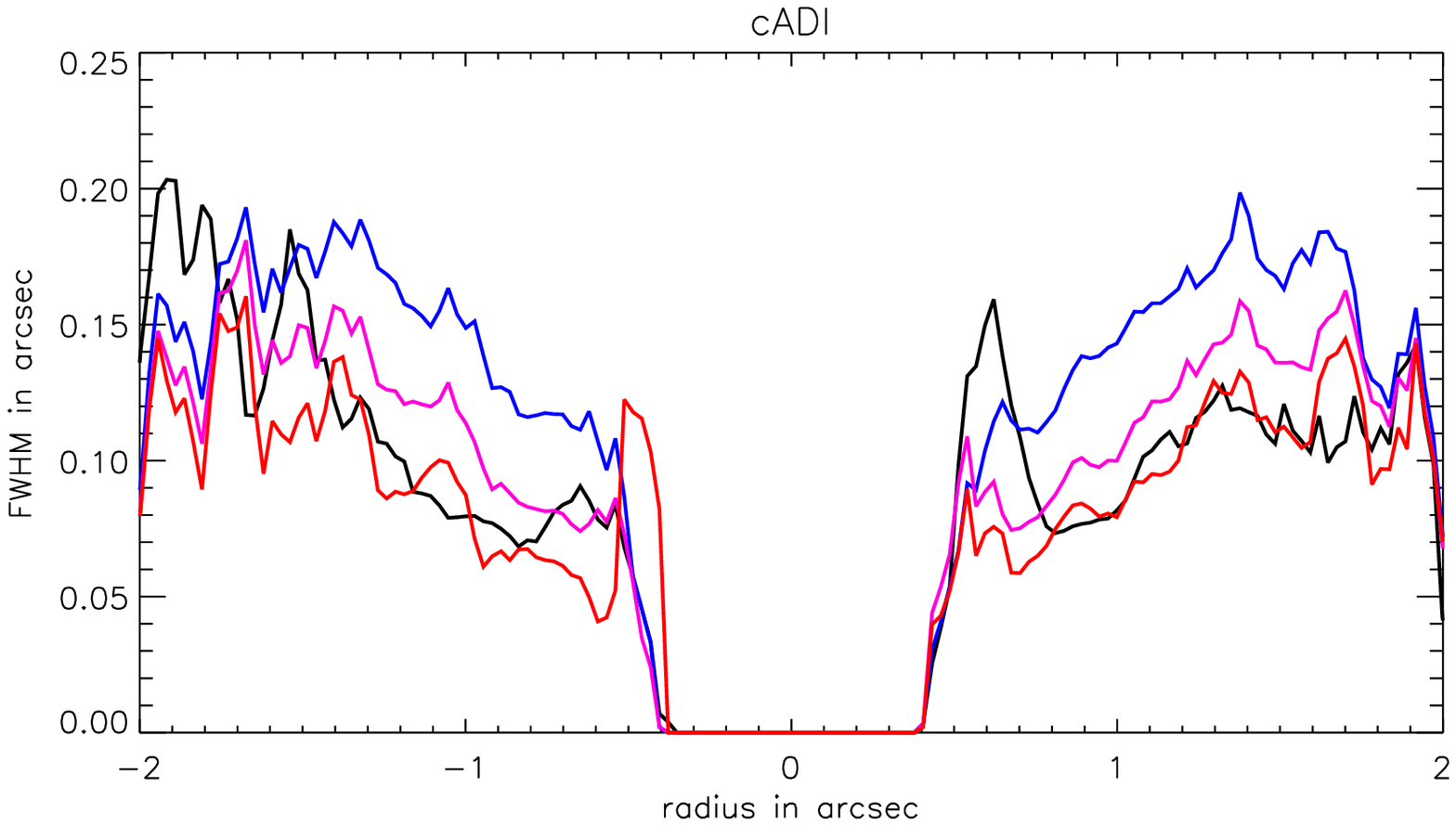}}
\centerline{\includegraphics[width=8.5cm]{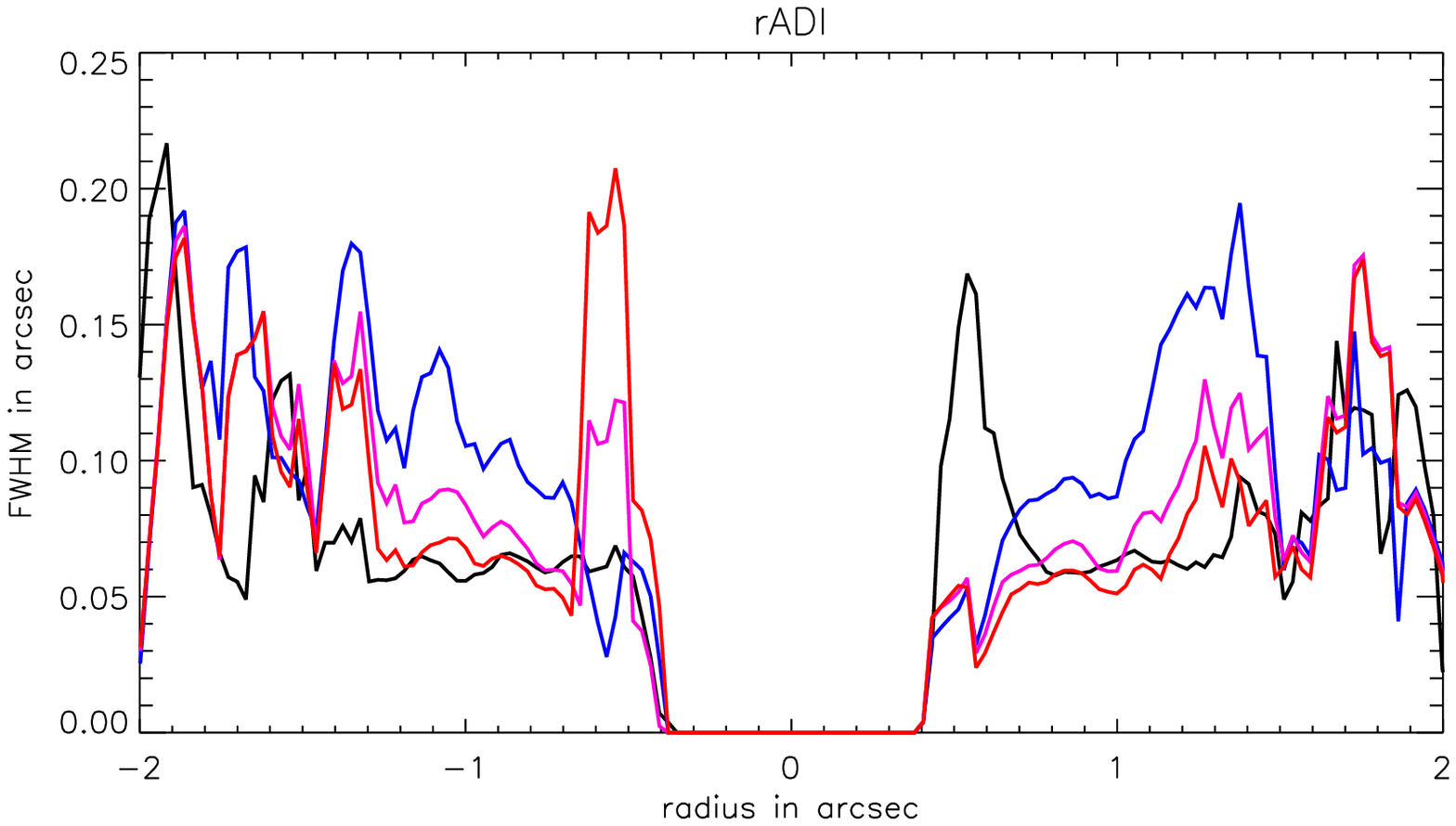}}
\centerline{\includegraphics[width=8.5cm]{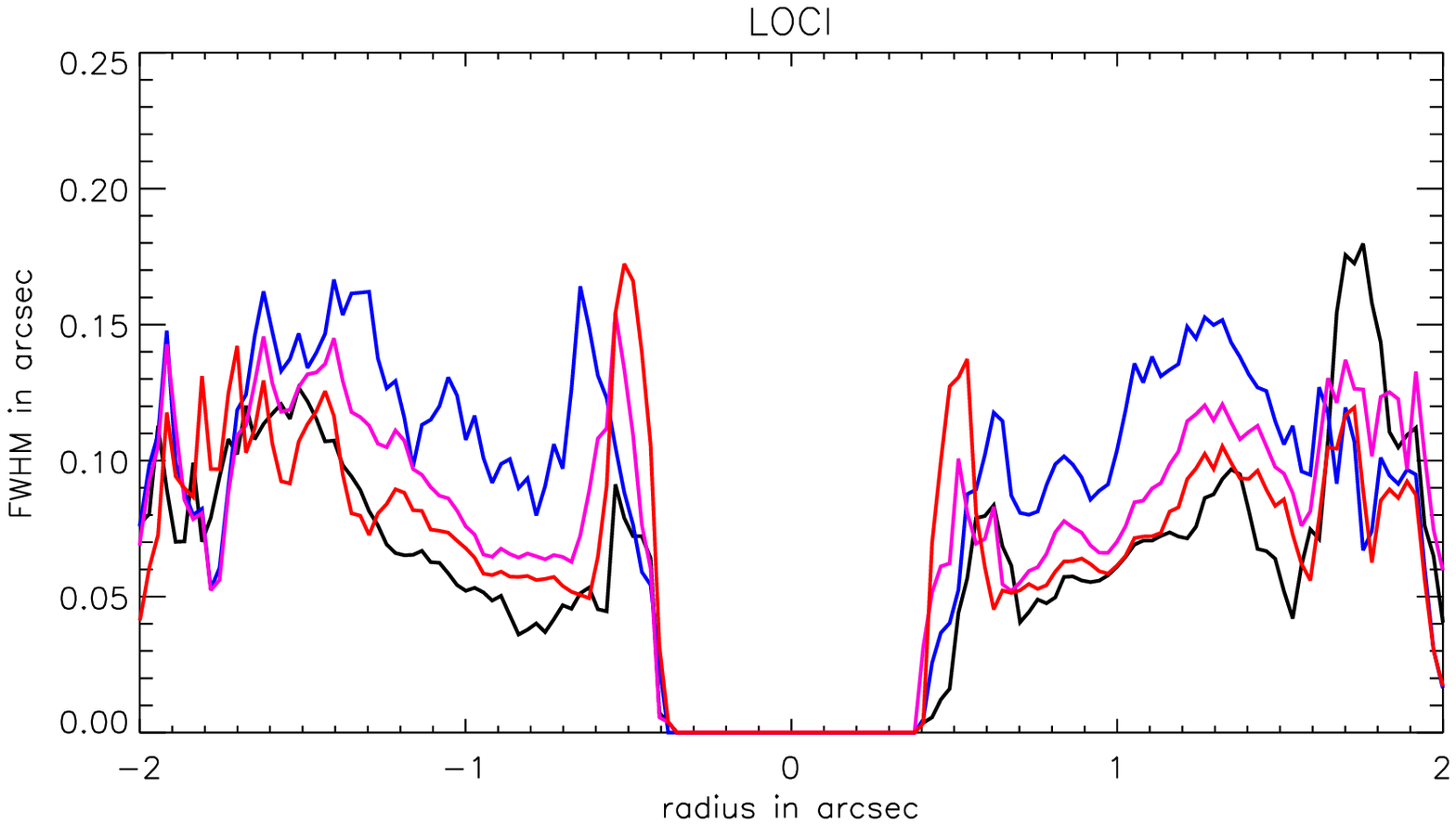}}
\caption{ Width of the observed disk (black line) in Ks for cADI (top), rADI (middle) and LOCI  (bottom) images, compared to the width of the model for $i=80\degb$ (blue line), $i=85\degb$ (magenta line) and $i=88\degb$ (red line). 
The NE side is at the left. }
\label{fig:fwhm}
\end{figure}

\subsection{Reliability of the spine deviation from midplane}
Following the observation reported in the previous section, our main objective was to test if the deviation of the midplane seen in Fig. \ref{fig:results} and Fig. \ref{fig:zoom} could be a bias produced by the ADI algorithms. Therefore, to overcome this issue and to transform observed measurements into physical ones requires calibration through models. In view of our images, unlike HST ones, we are not sensitive to the regions where the disk is posited to interact with the ISM (because of a lower sensitivity). Hence, we assumed a simple geometrical model of a ring-like dust disk. We used the GRaTer code \citep{Augereau1999, Lebreton2012} to calculate synthetic scattered light images of optically thin disks, assuming axial symmetry about their rotation axis, and inject such models in the real data. In this study, we assume a dust number density that writes (in cylindrical coordinates)

\begin{eqnarray}
n(r,z) = n_0
\sqrt{2}\left[\left(\frac{r}{r_0}\right)^{-2\alpha_{in}} +
  \left(\frac{r}{r_0}\right)^{-2\alpha_{out}}\right]^{-\frac{1}{2}}
\exp{\left(\left(\frac{-|z|}{H_0\left(\frac{r}{r_0}\right)^{\beta}}\right)^{\gamma}\right)}
\end{eqnarray}
where $n_0$ and $H_0$ are the midplane number density and the disk
scale height at the reference distance to the star $r=r_0$,
respectively. The slopes $\alpha_{in}$ and $\alpha_{out}$ are usually
chosen such that the density is rising with increasing distance up to
$r_0$ and is decreasing beyond ($\alpha_{in}\geq 0$ and
$\alpha_{out}\leq 0$). This is aimed to mimic a dust belt with a peak
density position close to $r_0$, a geometry appropriate for cold
debris disks { in isolation (no interaction with ISM)}. A ray-traced image is then calculated assuming a disk
inclination $i$ from pole-on viewing, and anisotropic scattering
modeled by an \citet{Henyey1941} phase function with asymmetry
parameter $g$ between $-1$ (pure backward scattering) and $1$ (forward
scattering). To limit the number of free parameters, we assumed a
gaussian vertical profile ($\gamma=2$) with a scale height
proportional to the distance to the star ($\beta=1$).
At this stage, we generated a limited number of models with some
assumptions on the remaining free parameters taken from the literature
\citep{Schneider2005, Moerchen2007, Debes2009}: $g=0$ (isotropic
scattering), $r_0=70$\,AU, $\alpha_{out}=-4$, $H_0=2$\,AU, $i = [80,
  85, 88]\degb$, $\alpha_{in}=[0,2]$. The vertical height $H_0$ is
consistent with the natural thickening of debris disks as defined by
\citet{Thebault2009}, although chosen close to the lower boundary of
the proposed range in that paper. 
Hence in our case, the aspect ratio is assumed constant:  $H/r \simeq 0.029$.
We note that the simplified version of GRaTer used in this study does not handle perfect edge-on viewing i=90$\degb$.

\begin{figure}[t]
\centerline{\includegraphics[width=8.5cm]{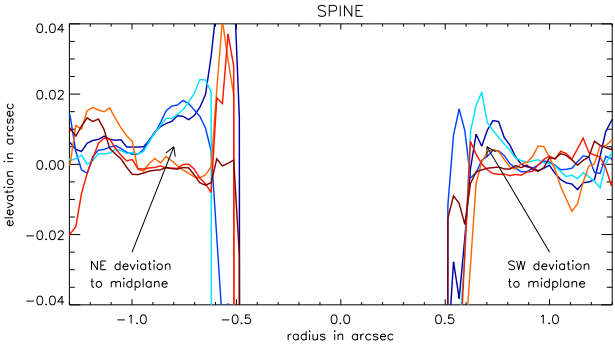}}
\caption{ Spine of the disk in Ks for the observed disk (bluish lines), and the $i=88\degb$ model (reddish lines) as measured in cADI, rADI  and LOCI. The elevation from the midplane is measured for $PA = 47.4\degb$. Arrows indicate where the deviation to the midplane is observed. The NE side is at the left.}
\label{fig:spine}
\end{figure}

Images of the models are first convolved with the observed PSF { (obtained with the ND filter but no stop)}. While
the field rotates, the orientation of the disk changes with respect to
some PSF features (spiders and static phase aberrations). So
rigorously, a model must be first rotated at a given orientation and
then convolved with the PSF, and the convolution is repeated for each
parallactic angle in the data cube. We checked however that our
conclusions are not affected if the convolution is applied once and the convolved model is rotated. 
Likely, the AO-correction quality is not good enough in this particular case
 to make a difference. We adopt this faster solution to save computing time.
Model disks were inserted at 90$\degb$ from the observed disk so there is
no possible overlap in between. A similar strategy was adopted by
\citet{Thalmann2011}. 
We used the same approach as described above  for measuring the $PA$.
The fitting of the { observed} and { model} disk vertical
profiles versus radius yields the disk spine position, vertical
thickness and intensity. Vertical thickness appears to be a good criterion to disentangle
between several disk inclinations ($i$ parameter). Figure
\ref{fig:fwhm} plots the vertical thickness (FWHM of a gaussian) measured on the observed disk
(black line) and the model disks (colored lines) for the three inclinations and
the three algorithms.  Qualitatively, we found that} the best matching with the models 
is achieved for an inclination of $i=88\degb$. This will be confirmed in the next section
with a more precise model fitting but already seems different than 
the values of \citet{Schneider2005} where $77\degb \leq i \leq
82\degb$. Also, we note that the shape of the FWHM profiles as a
function of radius is quite well reproduced with the model, providing support to
a constant disk aspect ratio. This suggests that the disk is opened
beyond about 1". Finally, we measure for this particular inclination
of 88$\degb$, a minimal observed vertical thickness for the observed and
the { model} disks of about 0.07" at a distance of about 0.85"
(96\,AU) in cADI images. This is consistent with the value of $H_0$ used in the model given the effect due to the inclination and the convolution with the PSF.

Figure \ref{fig:spine} shows for the most likely inclination ($i = 88\degb$) 
the disk spine position for cADI, rADI and LOCI images in Ks and for both the observed and model disks. 
The deviation to the midplane is evidenced especially in the NE side for Ks images between 0.6 and 1" and reaches at maximum $0.025"$ ($\sim$1 pixel).
On average the deviation has an amplitude of about 0.015" and spans across 15 pixels (which is about 6 resolution elements). The noise level per resolution element is equivalent to about 0.005".
Therefore, the deviation from the midplane is significant at the $\sim$7.5-sigma level.
 A low-significance symmetrical deviation is barely seen in the SW side too with the same amplitude. In H-band images, the deviation is more difficult to detect. It is still visible in the NE side in rADI and LOCI images only. 
There is also a possible deviation at larger radii ($>1.2"$) on both sides and towards the North-West which may comply with the change of $PA$ noted by \citet{Debes2009} but our data lack of sensitivity compared to HST in these regions for a confirmation.
We checked that other models mostly lead to the same conclusion. In addition, we found that the deviation in the NE side at 0.6-1" is resilient to a variation of the $PA$ within the error bar (0.3$\degb$). 

Finally, as the LOCI performs locally the injection of model disks at 90$\degb$ from the observed disk, the procedure might be inappropriate to calibrate the bias of the algorithm. 
Instead, in the LOCI images of Fig. \ref{fig:results}, we calculated the coefficients of the linear combinations for each zone and apply them to a model disk at the same PA than the observed disk (but in an empty data cube) and still observe a nearly flat midplane of the model disk in the 0.6-1" region. 
To summarize, an inclined model disk with a symmetrical vertical intensity distribution with respect to the midplane ($g=0$, $i<90\degb$) produces the same symmetry in the ADI processed images.
Therefore, the shape of the midplane observed in Figs. \ref{fig:results}, \ref{fig:zoom} and \ref{fig:spine} is definitely not due to a bias of the ADI process but is real. 
Interestingly and as an additional evidence, \citet{Esposito2012} images from Keck reveal (at least qualitatively) the same deviation from the midplane at the same separations.

\begin{table}[t]
\caption{Slopes { (as fitted to power laws)} of the different components in the surface brightness of cADI images. }
\begin{center}
\begin{tabular}{lcccc} \hline \hline
			&	\multicolumn{2}{c}{H band}			&	\multicolumn{2}{c}{K band}	\\
			&	NE		&	SW		& 	NE		&	SW						\\ \hline
inner slopes	&	1.9		&	1.4		&	-0.6		&	0.0						\\
outer slopes	&	-3.4		&	-3.2	 	&	-4.6		&	-3.9					\\ \hline
\end{tabular}
\end{center}
\label{tab:slopes}	
\end{table}%

\section{Photometric analysis}

\begin{figure}[t]
\centerline{\includegraphics[width=8.5cm]{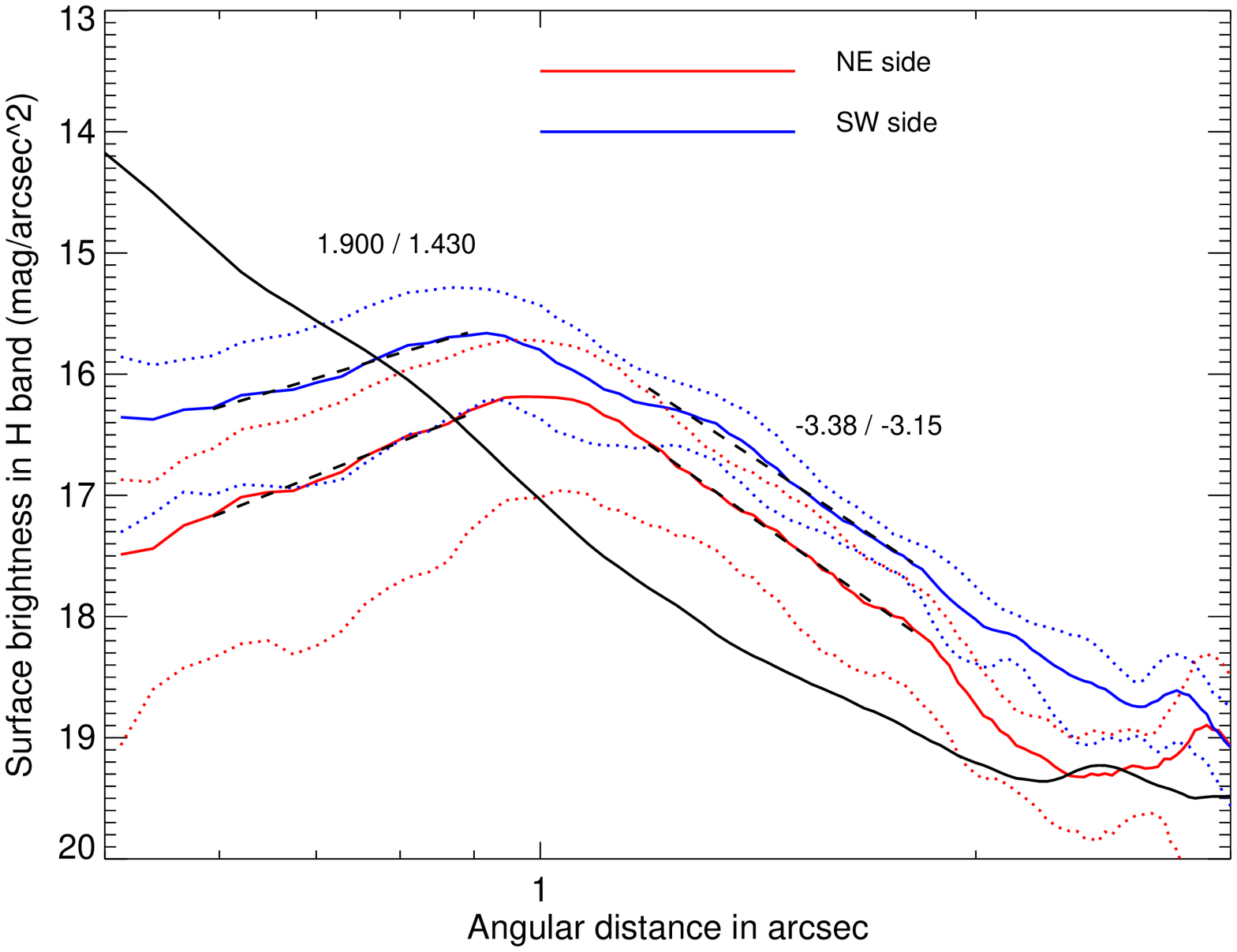}}
\centerline{\includegraphics[width=8.5cm]{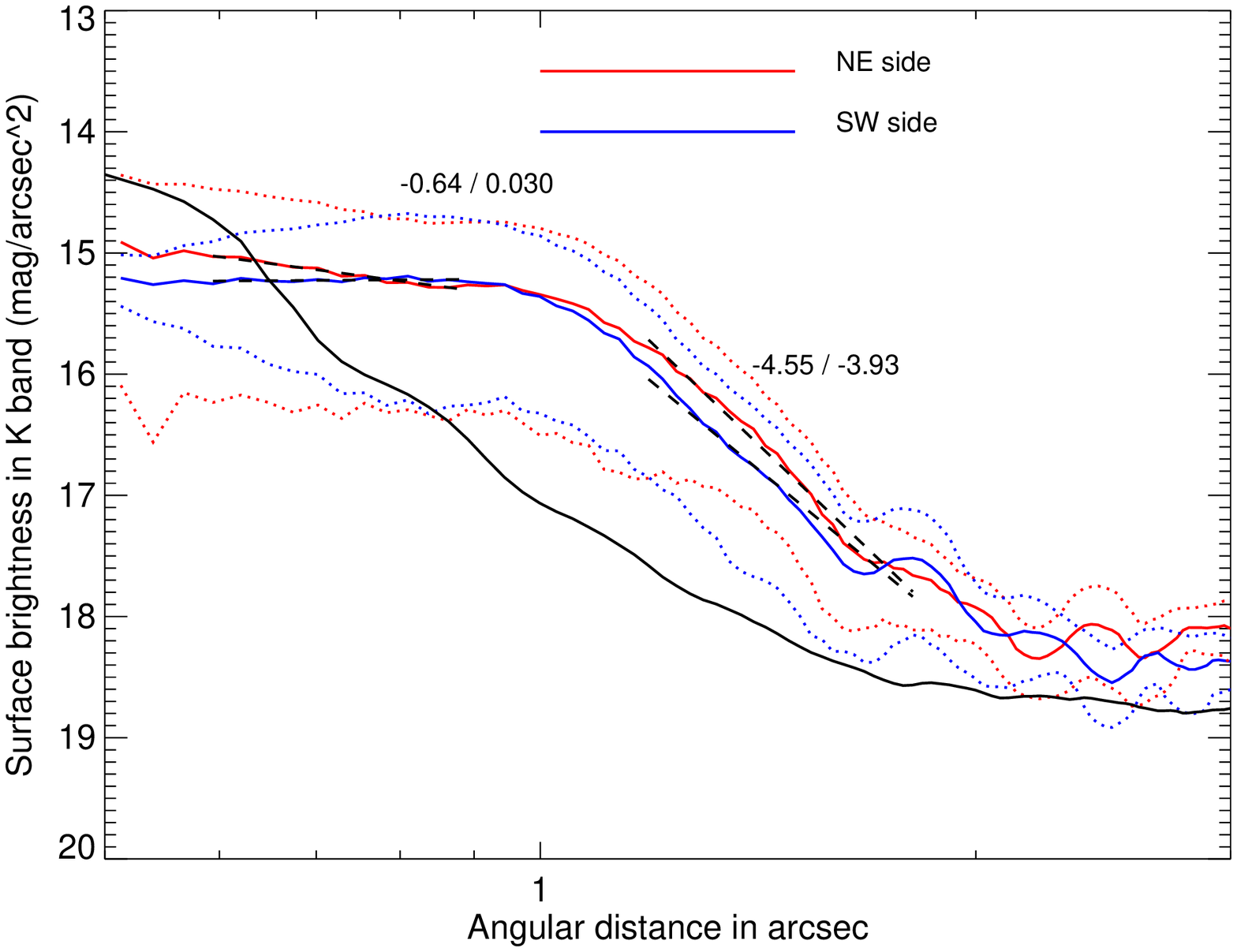}}
\caption{Surface Brightness expressed in mag/arcsec$^2$ as measured in the H band (top) and in the Ks band (bottom). Red and blue lines stand for the NE and SW sides respectively. Dotted lines show the disk intensity dispersion at 1 $\sigma$ within the slit window of 0.25". 
The black line indicates the noise level measured out of the slit window.
Dashed lines correspond to the fitted power law profile next to which the slopes are indicated. }
\label{fig:slopes}
\end{figure}

\label{sec:photom}
The photometric analysis of a disk observed in ADI can be problematic especially when the field rotation is small as in our case. The ADI effect can be understood in the resulting image as a transmission map projected on the sky in which the pattern depends on the disk brightness distribution.
It is for instance very interesting to note in \citet{Lagrange2012} that LOCI produces on the disk of $\beta$ Pic a very strong attenuation, just because it is very well resolved in the vertical dimension. 
The actual brightness distribution of the disk is therefore needed to properly account for this attenuation and derive reliable measures. 
This is the motivation for the modeling work in the next section. The Surface Brightness profiles presented in this section cannot be considered as the true photometry but are providing us with a starting point.  

As a first step, we performed the Surface Brightness (SB) measurement on cADI images  as cADI produces less self-subtraction than other algorithms and hence less photometric/astrometric biases. 
The cADI image of the disk is smoothed over 4 pixels, and isolated in a narrow slit of 0.25" in width aligned on the disk $PA$. From that, we extracted the mean intensity profile of the disk azimuthally averaged in angular sectors as well as the intensity dispersion and plot the results in Fig. \ref{fig:slopes}. 
Doing so and given the edge-on geometry, the intensity mean and dispersion are hence calculated in pixels located at same physical distances to the star (same flux received) but different elevations from the midplane.
In addition, we did the same measurement out of the slit to derive the noise level.
There is a clear break in the SB profiles near 1" (about 110\,AU) between a nearly flat  (or even positive) profile inwards and a negative slope outwards. The slopes given in Table \ref{tab:slopes} are measured at specific positions that we consider to be the most relevant: 0.6-0.9" for the inner slopes and 1.2-1.8" for the outer slopes.  
Clearly, the inner slopes have lower significance because 1/ they are measured across only 0.3", equivalent to 11 pixels  (4.5 resolution elements), and 2/ this region has higher speckle noise (as indicated by the noise level in Fig. \ref{fig:slopes}).  
Indeed, the error bar on the SB slopes can be large considering the intensity dispersion plotted in Fig. \ref{fig:slopes}. 
To be conservative we adopt an uncertainty of $\pm0.5$ for the index of the slopes. The inner slope is steeper in H than in Ks but 
this is not reliable given the noise level at such separations as seen in Fig. \ref{fig:results}. The outer slopes are very similar for the two sides of the disk (within error bars), but noticeably steeper in Ks than in H as already noted by \citet{Debes2009}. The disk also appears dimmer in H than in Ks which might be consistent again with HST data. As in the case of $\beta$ Pictoris we may expect that the disk of HD\,32297 contains a planetesimals belt at radii smaller than $\sim 110$\,AU which produces dust particles blown away from the system by the radiation pressure of the star \citep{Augereau2001}. Outer slopes close to -4 are compatible with such an assumption.

\section{Disk modeling}
\label{sec:model}

\subsection{Parameter space}

We now explore if the morphology of the disk can be accounted for by a model of scattered light images.
Using GRaTer \citep{Augereau1999}, we generated a grid of 576 models of which the parameters span relevant values (based on previous guess) for the disk of HD\,32297 as given below:
\begin{itemize}
\item[-]{inclination ($i$): 82, 85, 88, 89.5$\degb$ }
\item[-]{peak density position  ($r_0$): 80, 90, 100, 110, 120\,AU }
\item[-]{power law index of the outer density ($\alpha_{out}$): -4, -5, -6}
\item[-]{power law index of the inner density ($\alpha_{in}$): 2, 5, 10}
\item[-]{anisotropic scattering factor ($g$): 0, 0.25, 0.50, 0.75}
\end{itemize}

We considered only positive values of $g$ since we suspect that the
deviation from the midplane is in fact due to anisotropic scattering
towards the NW, but we cannot make any distinction with negative values and an inclination larger than 90$\degb$.
In addition to all possible combinations of disk
parameters we have two filters and three different images (cADI, rADI, LOCI)
for each, which totals 3456 prescriptions. 
Based on the analysis in section \ref{sec:morpho}, we rejected
an inclination of 80$\degb$ and $r_0 =70$\,AU as they are clearly not
compatible with our data. Similarly to section \ref{sec:morpho}, the
disk models were convolved with the PSF, injected at 90$\degb$ from
the disk PA and normalized with the observed disk in the outer part of the
disk (40-50 pixels for Ks and 30-40 pixels for H). However, the
intensity normalization is not extremely accurate for two reasons:
given the possible range of parameters the fake disk morphology can be
different than that of the observed disk, and also the ADI algorithms,
especially LOCI, produce a self-subtraction that depends on the
morphology.  Therefore, in our effort to match the observed disk to models
we proceeded in several steps. We first started with a global
minimization of images pixel to pixel and then we focused on
minimizing particular measurements
like the slopes in the Surface Brightness, and the spine of the disk. At the end, we are looking for a set of disk parameters [$i$, $r_0$, $\alpha_{out}$, $\alpha_{in}$, $g$] that matches as far as possible the observed disk. 

\subsection{Global minimization}

As for the global minimization we isolated the relevant part of the disk with a slit centered on the disk PA respectively PA+90$\degb$ for the model, 0.5" in width for Ks respectively 0.3" in H and restrained the radius to the range 0.65-2" for H and Ks. 
 These numbers were carefully chosen to retain only the pixels containing flux from the observed disk
 and hence optimize the minimization. Then for each model we used a standard $\chi^2$ metric as follow: 

\begin{equation}
\chi^2 =  \sum_{ij}\frac{\left(I_{disk}(i,j)-I_{model}(i,j)\right)^2}{\sigma_{disk}^2(i,j)}
\end{equation}

where $I_{disk}$ and $I_{model}$ are the images of the observed disk and that of the model for the retained pixels $(i,j)$, and $\sigma^2_{disk}$ is derived from the azimuthal dispersion of $I_{disk}$ to account for the errors.
As we run the minimization in both sides NE/SW of the disk simultaneously, it implicitly assumes that the disk is symmetrical about the minor axis. 
The $\chi^2$ minimum usually gives the adequate model that best matches the data, but to prevent degeneracies between parameters of the model we selected instead the 10 lowest minima corresponding to 10 sets of model parameters per ADI algorithm. Combining the three algorithms amounts to 30 sets of parameters. 
It would not be appropriate to derive averaged values nor error bars since our grid of models is discrete and cannot be interpolated. 
Instead, we searched independently for each parameter among these sets which value has the highest occurrence as a measure of the significance (value with the highest occurrence/30). This quantity is a number between 0 and 1 and does not convey an assumption of the error bar.
The results are given in Table \ref{tab:min_global} for both H and Ks. 
In the particular case of $r_0$ we found that both 110 and 120\,AU were equally probable in the Ks image.
The significance (indicated in parenthesis in Table \ref{tab:min_global}) has to be larger than the inverse of the number of allowed values for each parameter.
Some parameters are very well constrained like $i$ and $g$ (at least in Ks) supporting a quite large inclination of 88$\degb$. 
The value for $r_0$ is in very good agreement with the slope break in the Surface Brightness measured by \citet{Debes2009} as in our measurement (110\,AU on average). Density profiles are more difficult to constrain especially $\alpha_{in}$ where both H and Ks are not consistent but we may expect chromatic dependence for this parameters or NE/SW differences. 
Finally, the global minimization clearly points to a noticeable anisotropic scattering of $g=0.5$. Minimization in H and Ks leads to the same values for $i$, $r_{0}$ and $g$ although the significance is always lower in H as a result of lower data quality. 

\begin{table}[t]
\caption{Best model parameters with the global minimization. Significance of values are indicated in { parenthesis}.}
\begin{center}
\begin{tabular}{cccccc}\hline\hline
filter		&	$i$ ($\degb$)	&	$\alpha_{out}$		&	$\alpha_{in}$	&	$r_0$ ($AU$)			&	$g$			   \\ 	\hline
H 		&	88 {\tiny(0.70)}	&	-6  {\tiny(0.43)}		&	10 {\tiny(0.47)}	&	110 {\tiny(0.47)}		&	0.50 {\tiny(0.37)}  \\
Ks 		&	88 {\tiny(0.97)}	&	-5 {\tiny(0.5)}		&	2 {\tiny(0.37)}	&	110, 120 {\tiny(0.5)}		&	0.50 {\tiny(0.90)}  \\ \hline
\end{tabular}
\end{center}
\label{tab:min_global}
\end{table}

\subsection{Minimization of Surface Brightness slopes and spine}

The global minimization approach is useful to get a global picture of the disk because it takes into account all the available information in all the relevant pixels and try to simultaneously adjust all the parameters of the model, assuming the disk is symmetrical with respect to the minor axis.  But since the intensity scaling of the model with respect to the observed disk can be problematic (as explained in the previous sub-section)
we tested alternative procedures. To refine the analysis, as a second step, we carry out the minimization with other criteria that are less dependent on pixel to pixel variations. Here, we set the inclination to $i=88\degb$ as a clear outcome of the global minimization analysis.

Starting with the Surface Brightness we proceed similarly to section \ref{sec:photom} to extract this quantity on both the observed disk and the model disks. We do not minimize the full Surface Brightness profile but just the slopes in the 4 areas of interest inward and outward of 1" and independently in the NE and SW sides. We used the same regions defined in section \ref{sec:photom} to minimize $\alpha_{out}$	and	$\alpha_{in}$, so respectively 1.2-1.8" and 0.6-0.9".
Again, the minimization provides 30 sets of parameters for each area and we searched for the highest occurrence. 
The results are shown in Table \ref{tab:min_slopes}.
The analysis of $\alpha_{out}$ shows a steeper index in the NE but differs for Ks and H but remains in agreement with the SB profiles of the observed disk plotted in Fig. \ref{fig:slopes}. 
The inner density indices are discrepant with the global minimization, although with larger significance and rather converge to an intermediate value of $\alpha_{in}=5$. 
The parameter $r_0$ is more subject to dispersion.  We concluded that these values are not representative since the range of $r_0$ falls unequally between the two regions where the minimization is performed. However, in this particular case, it is informative to use a mean instead of the highest occurrence since the allowed values for $r_0$ are better sampled than for the other parameter. This mean is much closer to 110\,AU in concordance with the global minimization.
Furthermore, the minimization of SB slopes is not appropriate to constrain the anisotropic scattering factor since the dispersion of solutions is too large ($0.00<g<0.75$). 

One last measurement was  carried out to further constrain the anisotropic scattering factor. We applied the procedure described in section \ref{sec:morpho} to evaluate the spine of the disk as we consider it can provide a { better estimation for the anisotropic scattering factor although potentially correlated to other parameters like the densities}.
For the same reasons as above, we expect these geometrical measurements to be less dependent on the intensity normalization of the fake disks. Results are reported in Table \ref{tab:min_spine} where we considered the NE and SW sides separately. 
Here, we did not expect significant results for  $\alpha_{out}$ and  $\alpha_{in}$ which are related to the disk intensity while we measure just coordinates for the spine. This is confirmed in Table \ref{tab:min_spine} where no particular value emerges.  
The radius $r_0$ shows a difference in between the NE (110-120\,AU) and SW (80-90\,AU) sides (although with a low significance in Ks) which actually is in agreement with the images where the NE side appear slightly more elongated than the SW side. 
\citet{Debes2009} also noted such a NE/SW asymmetry but in the opposite direction. 
However, they used directly the indication from the SB without modeling. Orbital motion of a hypothetical structure at such a separation is totally { inconsistent} between the two epochs. As the disk is seen almost edge-on, this asymmetry could be related to a { stellocentric} offset of the central cavity as in many other debris disks \citep{Boccaletti2003, Kalas2005b, Thalmann2011, Lagrange2012b} and then interpreted as a signpost of planet which is sculpting the disk. The relation between this potential offset and the anisotropic scattering remains to be explored. 
As for the factor $g$, the Ks band favors one single value of 0.25, while it is larger in H. We note that the largest value of the anisotropic scattering factor ($g=0.75$) is actually unrealistic according to our images since it would make the disk brighter than the stellar residuals in the zone $r<0.65"$. As we exclude this region dominated by the starlight in the minimization process, { the relevant pixels where large values of $g$ contributes are missed and hence} some large values may come out. It is surprising that the spine minimization yields a different value for $g$ than the global minimization (0.25 instead of 0.50) while it remains symmetrical. Nevertheless, the global minimization converges to $g=0.50$ with a much higher significance at least in Ks.  

\begin{table}[t]
\caption{Best model parameters with the Surface Brightness minimization. Significance of values are indicated in { parenthesis}.}
\begin{center}
\begin{tabular}{clcccc}\hline\hline
filter 			&	area					&	$\alpha_{out}$	&	$\alpha_{in}$	&	$r_0$ ($AU$)	&	$g$			\\ 	\hline	\multirow{4}{*}{H}	&	NE in				&	-			&	5 \tiny{(0.50)}	&	120 \tiny{(0.57)}	&	0.25 \tiny{(0.63)}\\	
				&	SW in				&	-			&	5 \tiny{(0.50)}	&	120 \tiny{(0.47)}	&	0.25 \tiny{(0.50)} \\
				&	NE out				&	-5  \tiny{(0.50)}	&	-			&	120  \tiny{(0.57)} 	&	0.75  \tiny{(0.33)} \\				
				&	SW out				&	-4  \tiny{(0.57)}	&	-			&	100	 \tiny{(0.40)}	&	0.25  \tiny{(0.30)} \\ \hline					\multirow{4}{*}{Ks}	&	NE in				&	-			&	5 \tiny{(0.47)}	&	100 \tiny{(0.27)}	&	0.75 	\tiny{(0.30)}	\\	
				&	SW in				&	-			&	2 \tiny{(0.40)}	&	90 \tiny{(0.47)}		&	0.50	\tiny{(0.47)}\\				
				&	NE out				&	-6 \tiny{(0.70)}	&	-			&	120 \tiny{(0.30)}	&	0.75 \tiny{(0.73)} \\
				&	SW out				&	-5 \tiny{(0.47)}	&	-			&	100 \tiny{(0.37)}	&	0.00 \tiny{(0.47)}\\ \hline

\end{tabular}
\end{center}
\label{tab:min_slopes}
\end{table}

\begin{table}[t]
\caption{Best model parameters with the spine minimization. Significance of values are indicated in { parenthesis}.}
\begin{center}
\begin{tabular}{clcccc}\hline\hline
filter 				&	area		&	$\alpha_{out}$	&	$\alpha_{in}$	&	$r_0$ ($AU$)	&	$g$			\\ 	\hline
\multirow{2}{*}{H}	&	NE 		&	- 4 {\tiny(0.37)}			&	10 {\tiny(0.60)}		&	120 {\tiny(0.53)}	&	0.5 {\tiny(0.67)} \\
				&	SW 		&	- 6 {\tiny(0.73)}			&	5 {\tiny(0.37)}		&	80 {\tiny(0.43)}		&	0.75 {\tiny(0.70)} \\ \hline
\multirow{2}{*}{Ks}	&	NE		&	- 6 {\tiny(0.40)}			&	5 {\tiny(0.47)}		&	110 {\tiny(0.23)}	&	0.25 {\tiny(0.53)} \\	
				&	SW		&	- 4  {\tiny(0.53)}			&	2 {\tiny(0.57)}		&	90 {\tiny(0.33)}		&	0.25 {\tiny(0.67)} \\ \hline
				
\end{tabular}
\end{center}
\label{tab:min_spine}
\end{table}%

\begin{figure*}[t]
\centerline{
\includegraphics[width=6cm]{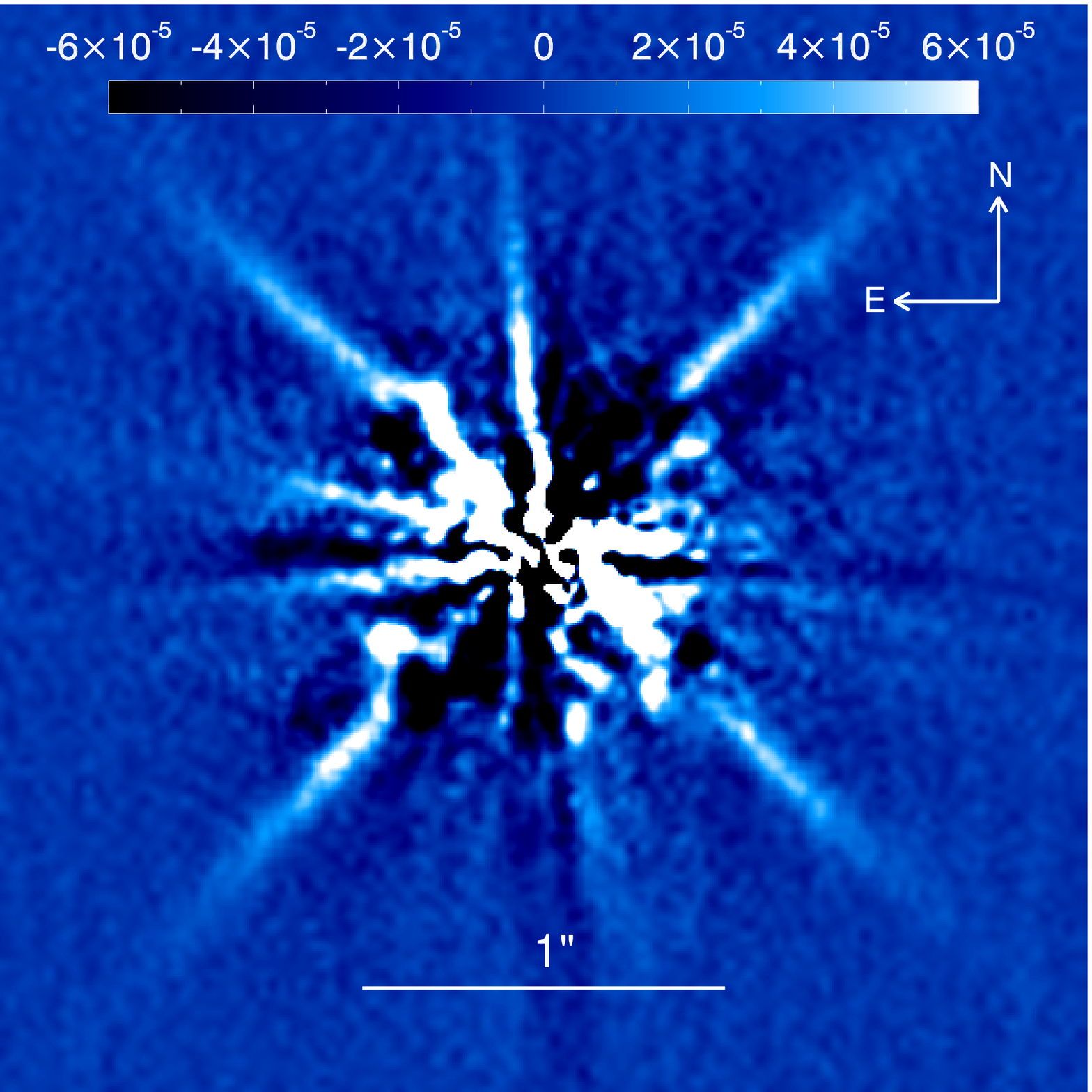}
\includegraphics[width=6cm]{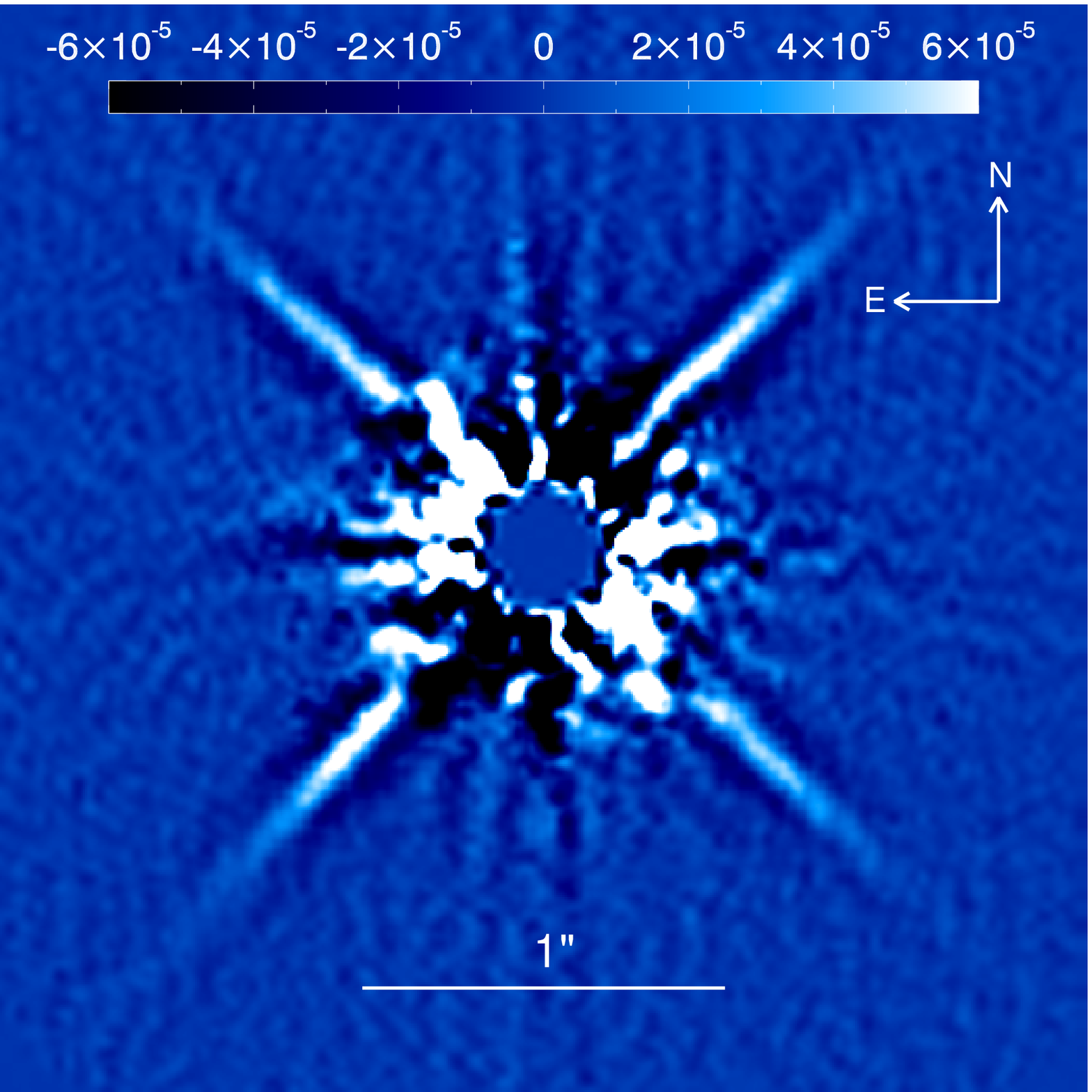}
\includegraphics[width=6cm]{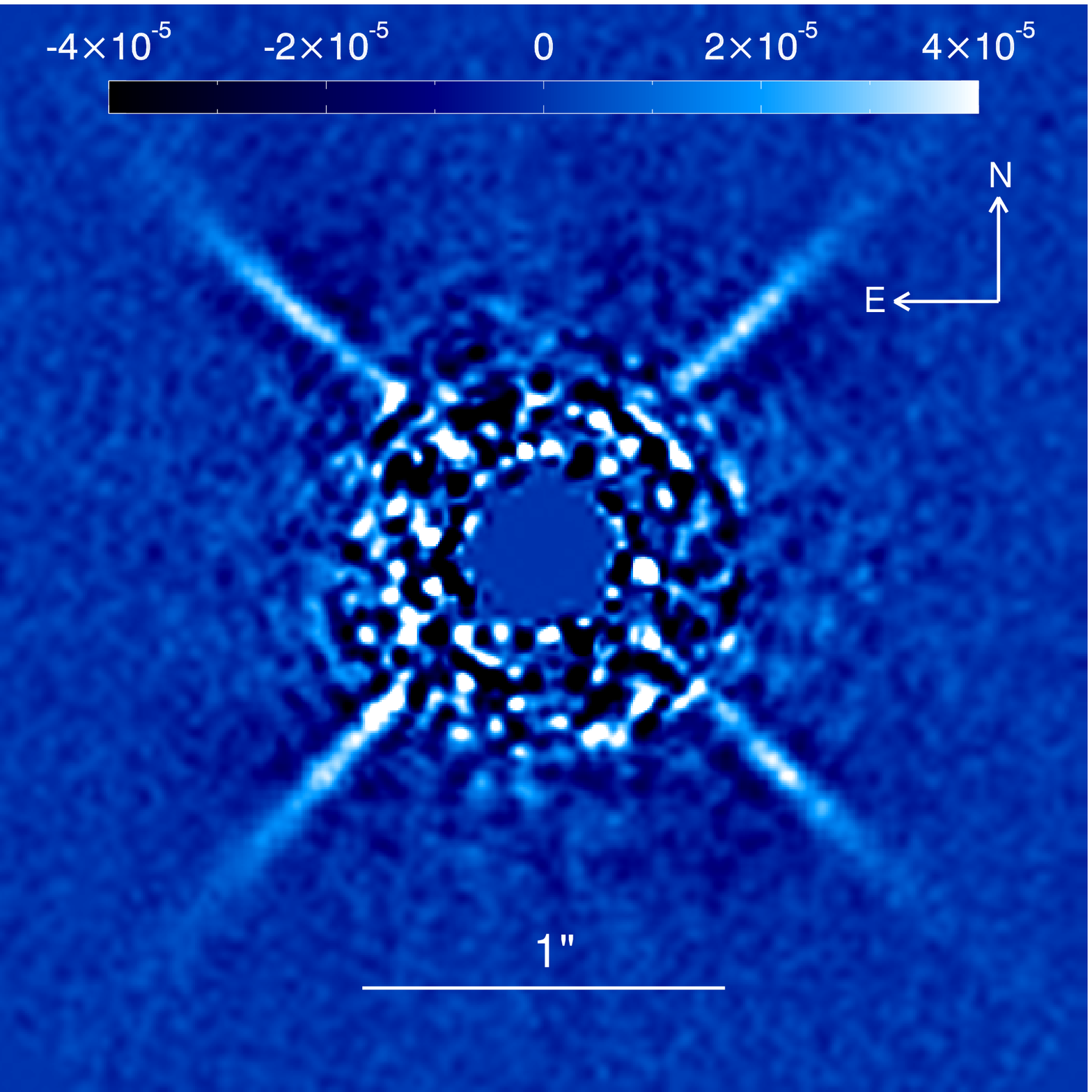}
}
\caption{Images of the best disk model (PA = 137.4$\degb$) in Ks compared to the observed disk (PA = 47.4$\degb$) with cADI, rADI, and LOCI  algorithms (left to right). 
The field of view of each sub-image is 3$\times$3''. Color bars indicate the contrast with respect to the star  { (not applicable to the display-saturated central residuals).}}
\label{fig:finalmodel}
\end{figure*}

\begin{figure}[ht]
\centerline{
\includegraphics[width=4.5cm]{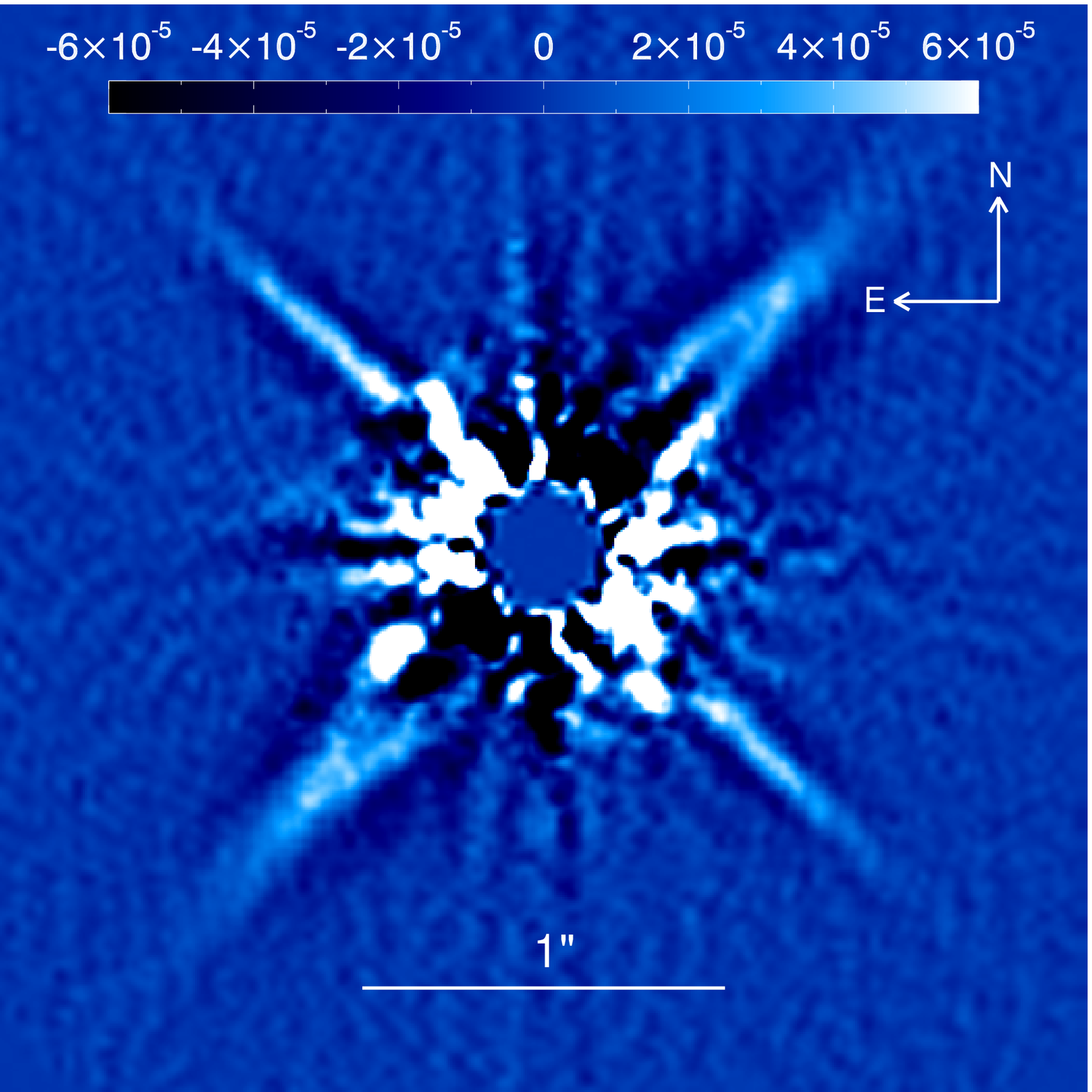}
\includegraphics[width=4.5cm]{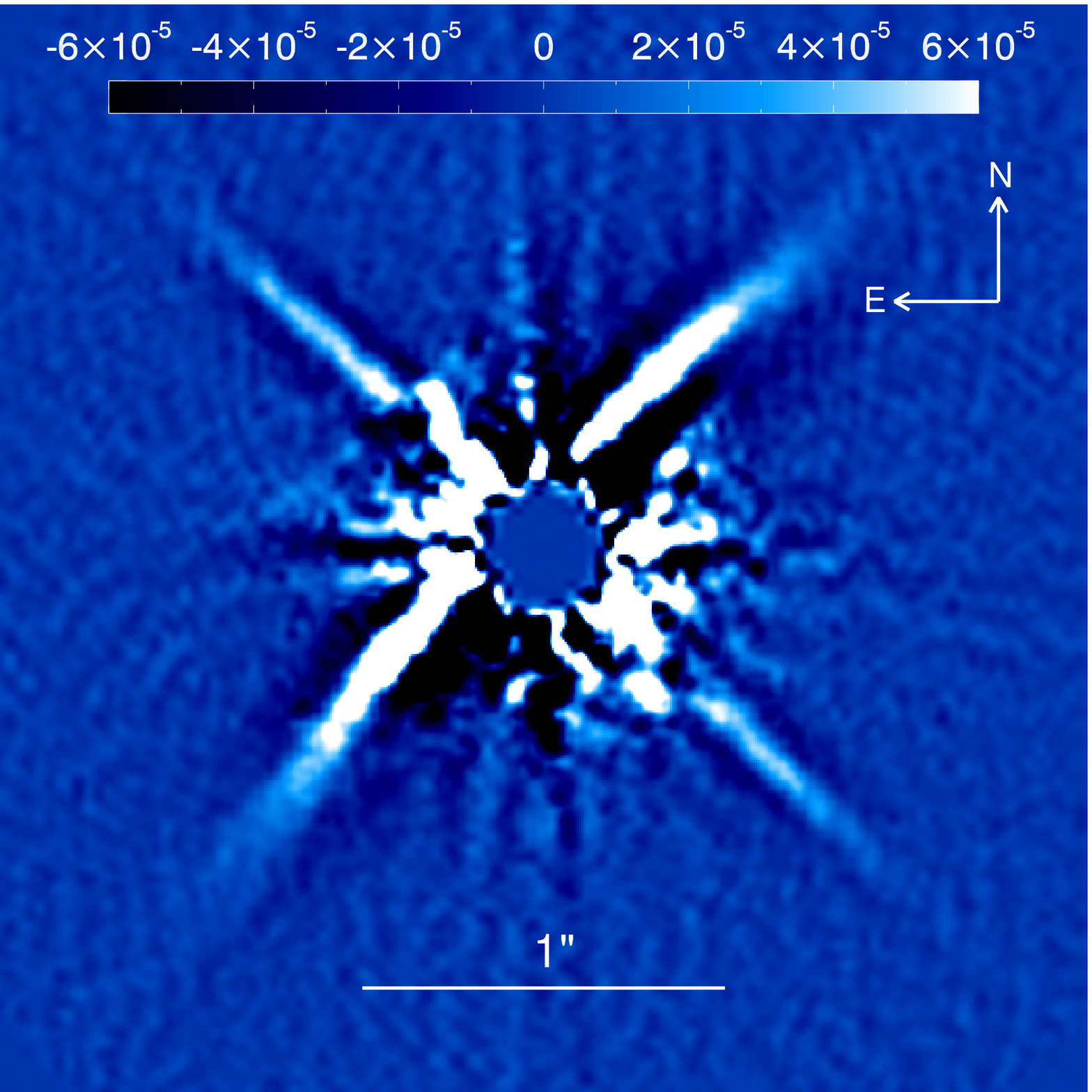}
}
\caption{Images of the disk model (PA = 137.4$\degb$) in Ks for $i=82\degb$ and $g=0$, $0.5$ (respectively left and right) compared to the observed disk  (PA = 47.4$\degb$) with rADI. The field of view of each sub-image is 3$\times$3''. Color bars indicate the contrast with respect to the star (not applicable to the display-saturated central residuals).}
\label{fig:model_inclination}
\end{figure}

\subsection{Discussion}

The determination of the model parameters is not perfect in the case of our data given the signal to noise ratio of the images while we also expect some degeneracies in between some parameters.
The disk of HD\,32297 being close to edge-on, { the inclination is definitely the easiest parameter to measure. We found an optimal value $i=88\degb$, larger than the original measure with HST which makes the disk more inclined than previously thought.  
This difference { of inclination is possibly explained and hence actually supports the scenario of}
interaction with the ISM \citep{Debes2009} which has noticeable effect in HST data { because of a better sensitivity to faint sources}}
Then, $r_0$ and $g$ are derived with quite good significance, at least for the global and spine minimization. 
For some parameters, it is reasonable to assume a higher weight from the images in Ks which are of much better quality than in the H band. 
Considering an axisymmetrical disk, we retain the values of the global minimization $r_0=110\,AU$ and $g=0.50$ as the optimal numbers to describe the disk morphology.
More difficult to constrain are the indices of the density profiles. First of all, inward from $r_0$ the number of pixels used to fit a slope is very much reduced. The 0.6-0.9" region used corresponds to about 11 pixels only in radius. Second, the noise introduced by the residual, { corresponding to} un-attenuated starlight produces bumpiness in the Surface Brightness which can easily affect the measurement of the slope exponent by 1, while it is the level of precision we are looking for { as imposed by the model parameter sampling}. Combining the global and Surface Brightness minimizations we choose  $\alpha_{out}=-5$ in Ks. In H, the values of $\alpha_{out}$ are discrepant. The SB minimization, in principle more accurate to measure the SB slopes, points to a lower value. 
In order to be consistent with the slopes shown in Fig. \ref{fig:slopes}, we adopt $\alpha_{out}=-5$ in H. This value is consistent with the predictions of \citet{Lecavelier1996} for a circular ring of planetesimals producing dust with collisions and expelled by the radiation pressure. It is also similar to the slopes measured in the debris disk around $\beta$ Pic \citep{Golimowski2006}.
Finally, the values of $\alpha_{in}$ are in practice clearly more difficult to measure but appear systematically larger in H than in Ks. To reproduce such a behavior we select $ \alpha_{in}=5$ in H and   $ \alpha_{in}=2$ in Ks.

Table \ref{tab:min_combined} gives the final choice of parameters combining the results presented above in Tables \ref{tab:min_global}, \ref{tab:min_slopes} and \ref{tab:min_spine}. As for illustration, a { model} disk with the parameters for the Ks band is shown in Fig. \ref{fig:finalmodel} at 90$\degb$ from the observed disk.
In both filters, the parameters that describe the morphology of the disk as projected on the sky ($i$, $r_0$ and $g$) are strictly identical. This is definitely a good indication that the minimization process is performing well. 
{ As an additional demonstration, Fig. \ref{fig:model_inclination} shows for the maximum inclination allowed by HST measurement \citep{Schneider2005} that the model would be discrepant with our disk image.
If the inclination were $i=82\degb$ with no anisotropy, we should see either the inner cavity or the a thicker disk depending on $ \alpha_{in}$. With an anisotropic scattering factor identical to our estimate ($g=0.5$), the deviation with respect to the midplane would be larger that what we measured.}
The parameters that rather depend on intensities ($\alpha_{out}$, $\alpha_{in}$) are on the contrary subject to questioning. The difference observed between the two filters can be due to either uncertainties or chromatic dependence. The accuracy of the data does not allow to firmly decide and these power low indices generate SB slopes which agree within error bars with those measured in Fig. \ref{fig:slopes}. However, we note that the depletion of the central cavity is not as steep as other cases like HD\,61005 \citep{Buenzli2010} or HR\,4796 \citep{Schneider1999, Lagrange2012b} as already mentioned by \citet{Moerchen2007} from mid IR images. 

\begin{table}[t]
\caption{Best model parameters retained.}
\begin{center}
\begin{tabular}{cccccc}\hline\hline
filter							&	$i$ ($\degb$)	&	$\alpha_{out}$	&	$\alpha_{in}$	&	$r_0$ ($AU$)	&	$g$	\\ \hline
H							&	88			&	-5			&	5			&	110			&	0.50 \\	
Ks							&	88			&	-5			&	2			&	110			&	0.50	\\ \hline
\end{tabular}
\end{center}
\label{tab:min_combined}
\end{table}%

\section{Surface Brightness of the best models}
\label{sec:photomcorrec}
The Surface Brightness profiles presented in Fig. \ref{fig:slopes} are impacted by the ADI process which produces a radial dependent self-subtraction of the disk. In particular, the parts of the disk that are closer to the star undergo a larger subtraction. 

\begin{figure}[t]
\centerline{\includegraphics[width=8.5cm]{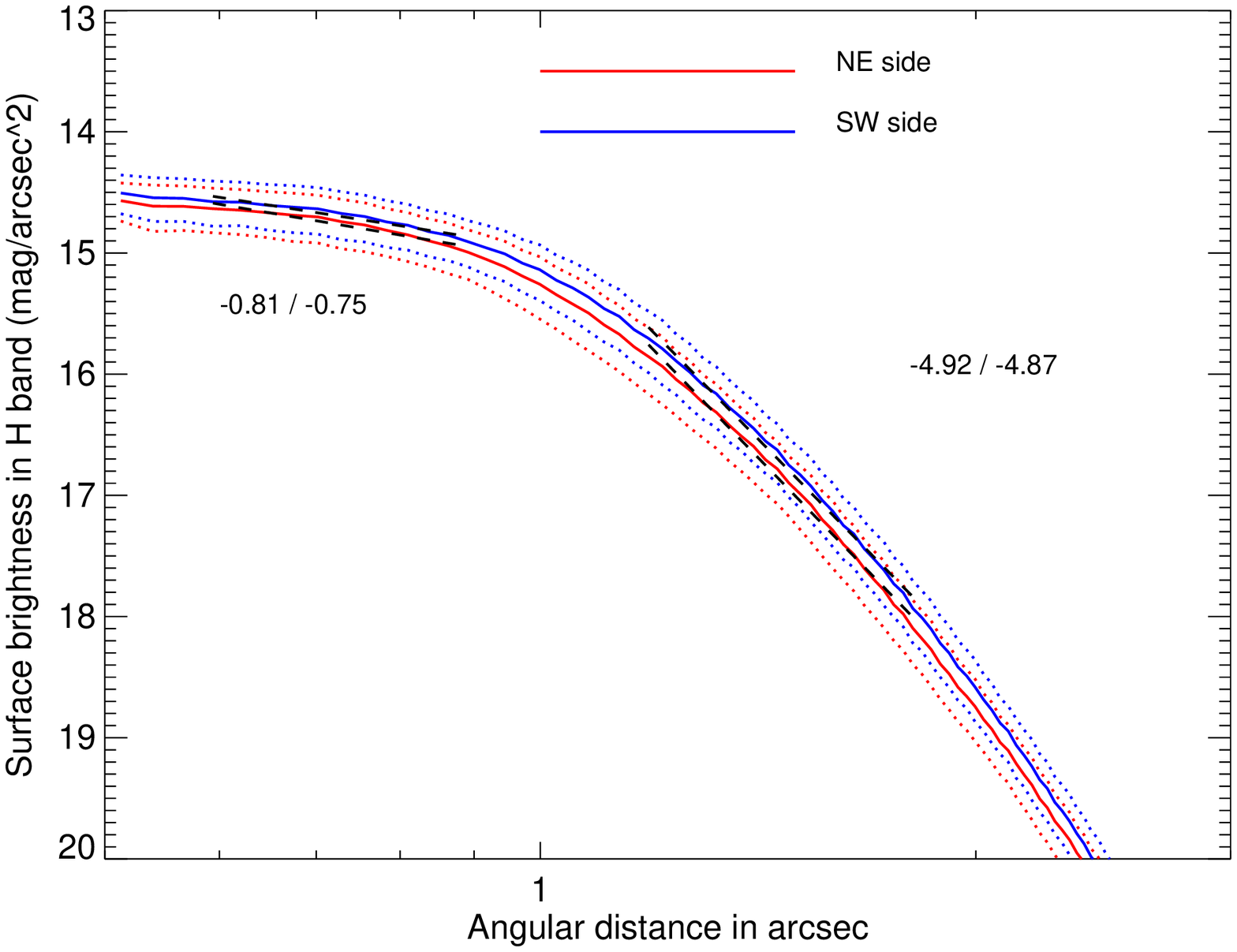}}
\centerline{\includegraphics[width=8.5cm]{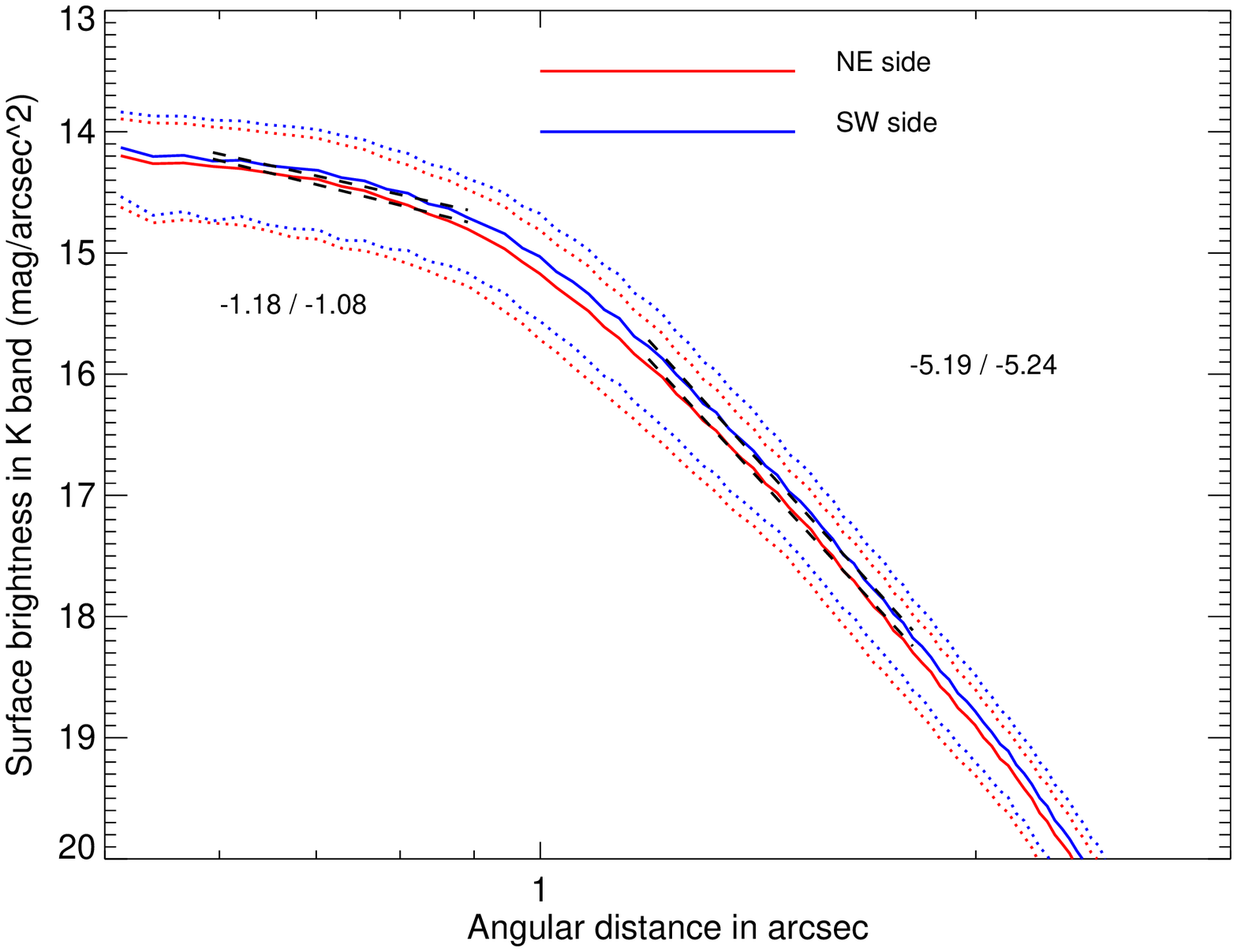}}
\caption{Surface Brightness expressed in mag/arcsec$^2$ as measured for the model with parameters $i=88\degb$, $r_0=110\,AU$, $\alpha_{out}=-5$, $\alpha_{in}=5$, $g=0.50$ (top), corresponding to the H band and for the parameters $i=88\degb$, $r_0=110\,AU$, $\alpha_{out}=-5$, $\alpha_{in}=2$, $g=0.50$, corresponding to the Ks band.
Red lines respectively blue lines stand for the NE side respectively the SW side. 
Dotted lines show the disk intensity dispersion at 1 $\sigma$ within the slit window of 0.25". 
Dashed lines correspond to the fitted power law profile next to which the slopes are indicated. }
\label{fig:slopescorrected}
\end{figure}

However, since the best model has been identified we can attempt to recover the most likely SB profile corrected from the self-subtraction by measuring it directly on the model once normalized adequately. For that, we first compare the cADI image of the observed disk and the cADI image of the best model to refine the intensity ratio in between. We used this ratio to inject a fake disk in an empty data cube to mimic the field rotation. Instead of processing the data with one of the ADI algorithms we simply derotated the frames and sum them up. Then, we applied the same measurement as in section \ref{sec:photom}. The resulting SB profiles corresponding to the aforementioned parameters are shown in Fig. \ref{fig:slopescorrected}.
The SB profiles derived from the models are significantly different than those measured in cADI images. {\it  ADI has an obvious effect on the slopes of the SB as also noted in \citet{Lagrange2012b}}. The net result of the ADI with respect to the model, in our case, is to change the intensity variations along the midplane sometimes reinforcing the intensity gradient. 
Starting from a negative inner slope in the models (Fig. \ref{fig:slopescorrected}) it becomes nearly flat (in Ks) or even positive (in H) in the ADI images (Fig. \ref{fig:slopes}). Similarly, the outer slope exponents are increased by $\sim 1$. We were not able to reproduce the steepness of the inner profile in H but the negative areas in the images (Fig. \ref{fig:results}) suggest a stronger effect of the reference frame subtraction than in Ks \footnote{ In the H band, the decline close to the axis of the SB as observed in Fig. \ref{fig:slopes} is caused by the negative pattern that appear in Fig. \ref{fig:results} (bottom left) at a different $PA$ than the disk but still affects the disk. It is due to a poor estimation of the reference image essentially because of the diffraction pattern decorrelation over time. We suspect that this butterfly pattern (positive and negative) is due to the secondary spiders. The angle is very close to the 109$\degb$ that separates two VLT spiders. This residual pattern is seen in several other NACO images.}. 
Finally, the profiles for the models are brighter as expected since the ADI self-subtraction is accounted for. The cADI image in Ks is about 0.3 mag/arcsec$^2$ fainter than the model at 1" (corresponding to $r_0$). In H, the attenuation can be as large as $\sim$1 mag/arcsec$^2$ at the same radius. 

\section{Limit of detection to point sources}
\begin{figure}[t]
\centerline{
\includegraphics[width=8.5cm]{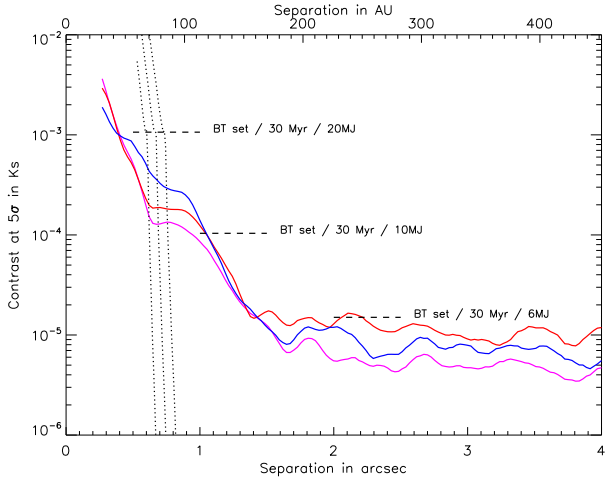}
}
\caption{Detection limit to point sources at 5$\sigma$ for cADI ({ magenta}), rADI (red) and LOCI (blue). The expected intensity of some planetary objects from the BT-SETTL model are shown for an age of 30 Myr { (dashed lines). The dotted lines correspond to the mass/separation range of planets able to sculpt the disk and assuming that the inner edge of the ring is located at 90, 100 and 110\,AU. }}
\label{fig:limdet}
\end{figure}

Our data feature the highest contrast ever reported for HD\,32297 in the Ks band near $\sim$1" hence suitable to derive limit of detection of point sources in the planet/brown dwarf mass regime. In this particular case, it is relevant to measure the contrast along the midplane as the orbits of hypothetical planets will be likely aligned with the disk plane. 
For each image in Fig. \ref{fig:results} the disk is isolated in a 0.3" slit window centered on the midplane. The standard deviation is calculated in the pixels that are vertical to the midplane at each radii (we check that the result do not strongly depends on that slit width).
The photometric biases introduced by the self-subtraction of the three ADI algorithms were calibrated with nine fake planets (copy of the PSF) injected into the data cube and located at specific locations (from 0.25" to 5" and 90$\degb$ from the midplane  to avoid confusion with the disk). An attenuation profile is interpolated on these flux losses assuming that it varies monotonically with radius (not necessary true for LOCI) and used to correct for photometric biases and to provide calibrated contrast curves.
 The 5$\sigma$ contrast curves are presented in Fig. \ref{fig:limdet}. As it is measured in the disk midplane instead of the background noise, the limit of detection is similar for the three algorithms.
We plot also the level of contrast for three planet masses: 6, 10 and 20 M$_\mathrm{Jupiter}$ assuming an age of 30 Myr \citep{Kalas2005} and considering the BT-SETTL models from \citet{Allard2011}. However, we note that the age of HD\,32297 is not very well determined. We can definitely reject the presence of { brown dwarfs} at separations larger $\sim$50\,AU { and closer than $\sim$600\,AU (FOV limitation)}. Due to the distance of the star (113\,pc) it is impossible to put any constraints on planetary mass objects at closer separations. At 110\,AU, the distance of the inner ring, the detection limit reaches $\sim$10M$_\mathrm{Jupiter}$. 
Assuming the inner cavity is sculpted by a planet and following the approach presented in \citet{Lagrange2012b} we can estimate the position of that hypothetical planet. According to \citet{Wisdom1980} { and \citet[][Eq. 23]{Duncan1989}} the distance of the planet to the rim of the ring ($\delta a$) is related to the planet/star mass ratio ($M_p/M_s$) with the relation:
\begin{equation}
\delta a/a = 1.3 \times (M_p/M_s)^{2/7}
\end{equation}  
where $a$ is the semi-major axis. So $a+\delta a$ is the distance of the inner rim which is lower than $r_0$.
{ The same formulation is used by \citet{Chiang2009} in the case of Fomalhaut b.}
We calculated $M_p$ for several assumptions on $a+\delta a$ (90, 100, 110\,AU) and taking  $M_s=2.09 M_\odot$. 
We interpolated these masses on the BT-SETTL model,  hence converted into absolute magnitudes and into contrasts. 
Smaller values of $a+\delta a$ would lead to unrealistic masses in the stellar domain. The dotted lines in Fig. \ref{fig:limdet} show the combination of planet masses and separations that are able to produce an inner ring at 110\,AU. The comparison between dynamical constraints (dotted lines) and detection limits (solid lines) indicates a possible range of separations and masses for the planet: a $> 65\sim85$\,AU and $M_p<$12 M$_\mathrm{Jupiter}$. 
We stress that this estimation is only valid for a planet at the quadrature (projected separation = physical separation). However, given the steepness of the dynamical constraints, if a planet produces the ring it is necessarily close to the rim or it would have been massive and hence detected. 

\section{Summary \& conclusions}
We { present} high contrast images of the debris disk around HD\,32297 obtained with NACO at the VLT in the near IR H and Ks bands. To improve the attenuation of the starlight and the speckle rejection we combined phase mask coronagraphy and ADI { algorithms}. The disk { nearly edge-on} is detected  in the two filters { at radial distances} from 0.65" to 2" (74 to 226 \,AU at the distance of 113\,pc) owing to various ADI algorithms. 
ADI is a very efficient method to achieve detection but induces some { photometric} biases that have to be { corrected for
in  the data analysis for a proper interpretation}. 
We observed a deviation from the midplane near 0.8" in the NE side that is clearly not an ADI bias. We focused our effort on the morphological modeling of the disk using a numerical code GRaTer, which produces synthetic scattered light images of debris disks. Following previous observations, we considered a disk containing a planetesimal belt with dust distribution inward and outward. We generated a 5-parameter grid of models and applied several sorts of minimizations to constrain the disk morphology. 

As a first result, we found { in Ks and H bands} a higher inclination than previously published for $1.1\mu m$ (88$\degb$ instead of 80$\degb$). In our images, the vertical extent of the disk together with the disk parameters found from modeling are definitely not consistent with such a disk inclination. However, we suspect that the isophotal ellipse fitting used in former data \citep{Schneider2005} can be biased by the strong interaction with the interstellar medium as revealed by \citet{Debes2009}, especially when considering the lower angular resolution and conversely the higher sensitivity to faint structures of NICMOS/HST as opposed to NACO.
In addition, we did not found any significant NE/SW brightness asymmetry like in \citet{Schneider2005} and  \citet{Mawet2009} although we are { also observing at near IR} wavelengths { too}. 

Furthermore, the minimization of the disk parameters allowed to constrain the radius of the planetesimal belt at 110\,AU. A NE/SW asymmetry is suspected, possibly due to an offset of the planetesimal belt with respect to the star. We showed that the deviation from the midplane is { likely} the result of anisotropic scattering. One edge of the disk is therefore brighter. Whether it is linked to the properties of the dust or to the geometry (the hypothetical offset of the belt) and hence to a gravitational perturber is yet undetermined. It is remarkable that among the known debris disks, three of them surrounding A-type stars, namely, $\beta$ Pic, Fomalhaut, and HD\,32297 share the same position of the planetesimal belt near 100\,AU as a possible indication of a common architecture.

Intensity related parameters like the brightness slopes of the inner and outer parts are more difficult to constrain and the values we determined should be taken with caution as the uncertainty can be large. However, it gives an order of magnitude indicating that the outer parts have a power law index close to that expected for dust grains expelled from the system by radiation pressure. As for the inner parts, there is a significant difference between Ks and H band filters, but overall, the steepness is smaller than some other ring-like debris disks as already noted by \citet{Moerchen2007}.

Armed with a detailed morphological characterization it is now possible to start a dynamical study to understand the properties of the debris disk around HD\,32297 in regards of planetary formation or gravitational perturbation as it was performed in other cases \citep{Wyatt2005, Chiang2009}. Finally, the observations of debris disks from ground-based telescopes { in the near IR} are achieving comparable or even better angular resolution than from HST, owing to { adaptive optics. A similar statement can be made about contrast} owing to a combination of high contrast imaging techniques (coronagraphy, ADI). The next generation of extreme adaptive optics instruments, the so-called planet finders \citep{Beuzit2008, Macintosh2008, Hodapp2008}, will  not only revolutionize the search and characterization of young long-period planets but also the field of debris disk science pending the arrival of the James Webb Space Telescope \citep{Gardner2006}.  

\begin{acknowledgements}
We would like to thank G. Montagnier at ESO for conducting these observations and A. Lecavelier des Etangs for helpful discussion. We also thanks the referee for a very detailed and useful report. 
\end{acknowledgements}

\bibliography{hd32_bib.bib}
\end{document}